\documentclass[twocolumn]{aastex63}
\pdfoutput=1
\usepackage{xcolor}
\usepackage{makecell}
\usepackage{pbox}

\received{\today}
\submitjournal{ApJ}
\shorttitle{ASASSN-14ko}
\shortauthors{Payne et al.}

\def\gal{ESO 253$-$G003}

\begin{document}

\title{ASASSN-14ko is a Periodic Nuclear Transient in \gal{}}

\correspondingauthor{Anna~V.~Payne}
\email{avpayne@hawaii.edu}

\author[0000-0003-3490-3243]{Anna~V.~Payne}
\altaffiliation{NASA Fellow}
\affiliation{Institute for Astronomy, University of Hawai\`{}i at Manoa, 2680 Woodlawn Dr., Honolulu, HI 96822}

\author[0000-0003-4631-1149]{Benjamin~J.~Shappee}
\affiliation{Institute for Astronomy, University of Hawai\`{}i at Manoa, 2680 Woodlawn Dr., Honolulu, HI 96822}

\author[0000-0001-9668-2920]{Jason~T.~Hinkle}
\affiliation{Institute for Astronomy, University of Hawai\`{}i at Manoa, 2680 Woodlawn Dr., Honolulu, HI 96822}

\author[0000-0001-5661-7155]{Patrick~J.~Vallely}
\altaffiliation{NSF Graduate Research Fellow}
\affiliation{Department of Astronomy, The Ohio State University, 140 West 18th Avenue, Columbus, OH 43210, USA}

\author[0000-0001-6017-2961]{Christopher~S.~Kochanek}
\affiliation{Department of Astronomy, The Ohio State University, 140 West 18th Avenue, Columbus, OH 43210, USA}
\affiliation{Center for Cosmology and AstroParticle Physics, The Ohio State University, 191 W.\ Woodruff Ave., Columbus, OH 43210, USA}

\author[0000-0001-9206-3460]{Thomas~W.-S.~Holoien}
\altaffiliation{Carnegie Fellow}
\affiliation{The Observatories of the Carnegie Institution for Science, 813 Santa Barbara St., Pasadena, CA 91101, USA}

\author[0000-0002-4449-9152]{Katie~Auchettl}
\affiliation{School of Physics, The University of Melbourne, Parkville, VIC 3010, Australia}
\affiliation{ARC Centre of Excellence for All Sky Astrophysics in 3 Dimensions (ASTRO 3D)}
\affiliation{Department of Astronomy and Astrophysics, University of California, Santa Cruz, CA 95064, USA}
\affiliation{DARK, Niels Bohr Institute, University of Copenhagen, Lyngbyvej 2, 2100 Copenhagen, Denmark}

\author{K.~Z.~Stanek}
\affiliation{Department of Astronomy, The Ohio State University, 140 West 18th Avenue, Columbus, OH 43210, USA}
\affiliation{Center for Cosmology and AstroParticle Physics, The Ohio State University, 191 W.\ Woodruff Ave., Columbus, OH 43210, USA}

\author[0000-0003-2377-9574]{Todd~A.~Thompson}
\affiliation{Department of Astronomy, The Ohio State University, 140 West 18th Avenue, Columbus, OH 43210, USA}
\affiliation{Center for Cosmology and AstroParticle Physics, The Ohio State University, 191 W.\ Woodruff Ave., Columbus, OH 43210, USA}

\author[0000-0001-7351-2531]{Jack~M.~M.~Neustadt}
\affiliation{Department of Astronomy, The Ohio State University, 140 West 18th Avenue, Columbus, OH 43210, USA}

\author[0000-0002-2471-8442]{Michael~A.~Tucker}
\altaffiliation{DOE CSGF Fellow}
\affiliation{Institute for Astronomy, University of Hawai\`{}i at Manoa, 2680 Woodlawn Dr., Honolulu, HI 96822}

\author{James~D.~Armstrong}
\affiliation{ Institute for Astronomy, University of Hawai\`{}i, 34 Ohia Ku St., Pukalani, HI 96768, USA}

\author{Joseph~Brimacombe}
\affiliation{Coral Towers Observatory, Cairns, QLD 4870, Australia}

\author{Paulo~Cacella}
\affiliation{Dogsheaven  Observatory,  SMPW  Q25CJ1 LT10B, Brasilia, Brazil}

\author{Robert~Cornect}
\affiliation{Moondyne Observatory, Bakers Hill, Western Australia, Australia}

\author{Larry~Denneau}
\affiliation{Institute for Astronomy, University of Hawai\`{}i at Manoa, 2680 Woodlawn Dr., Honolulu, HI 96822}

\author[0000-0002-9113-7162]{Michael~M.~Fausnaugh}
\affiliation{Department of Physics, and Kavli Institute for Astrophysics and Space Research, Massachusetts Institute of Technology, Cambridge, MA 02139, USA}

\author[0000-0002-1050-4056]{Heather~Flewelling}
\affiliation{Institute for Astronomy, University of Hawai\`{}i at Manoa, 2680 Woodlawn Dr., Honolulu, HI 96822}

\author[0000-0002-9961-3661]{Dirk~Grupe}
\affiliation{Department of Physics, Earth Science, and Space System Engineering, Morehead State University, 235 Martindale Dr, Morehead, KY 40351}

\author[0000-0003-3313-4921]{A.~N.~Heinze}
\affiliation{Institute for Astronomy, University of Hawai\`{}i at Manoa, 2680 Woodlawn Dr., Honolulu, HI 96822}

\author{Laura~A.~Lopez}
\affiliation{Department of Astronomy, The Ohio State University, 140 West 18th Avenue, Columbus, OH 43210, USA}
\affiliation{Center for Cosmology and AstroParticle Physics, The Ohio State University, 191 W.\ Woodruff Ave., Columbus, OH 43210, USA}

\author{Berto~Monard}
\affiliation{Bronberg Observatory, Center for Backyard Astrophysics Pretoria, PO Box 11426, Tiegerpoort 0056, South Africa; Kleinkaroo Observatory, Center for Backyard Astrophysics Kleinkaroo, Sint Helena 1B, PO Box 281, Calitzdorp 6660, South Africa}

\author[0000-0003-1072-2712]{Jose~L.~Prieto}
\affiliation{N\'ucleo de Astronom\'ia de la Facultad de Ingenier\'ia y Ciencias, Universidad Diego Portales, Av. Ej\'ercito 441, Santiago, Chile}
\affiliation{Millennium Institute of Astrophysics, Santiago, Chile}

\author[0000-0002-6294-5937]{Adam~C.~Schneider}
\affiliation{US Naval Observatory, Flagstaff Station, P.O. Box 1149, Flagstaff, AZ 86002, USA}
\affiliation{Department of Physics and Astronomy, George Mason University, MS3F3, 4400 University Drive, Fairfax, VA 22030, USA}

\author[0000-0003-3145-8682]{Scott~S.~Sheppard}
\affiliation{Carnegie Institution for Science, Earth and Planets Laboratory, 5241 Broad Branch Road, Washington, DC 20015, USA}

\author{John~L.~Tonry}
\affiliation{Institute for Astronomy, University of Hawai\`{}i at Manoa, 2680 Woodlawn Dr., Honolulu, HI 96822}

\author{Henry~Weiland}
\affiliation{Institute for Astronomy, University of Hawai\`{}i at Manoa, 2680 Woodlawn Dr., Honolulu, HI 96822}

\begin{abstract}

We present the discovery that ASASSN-14ko is a periodically flaring AGN at the center of the galaxy \gal{}. At the time of its discovery by the All-Sky Automated Survey for Supernovae (ASAS-SN), it was classified as a supernova close to the nucleus. The subsequent six years of $V$- and $g$-band ASAS-SN observations reveal that ASASSN-14ko has nuclear flares occurring at regular intervals. The seventeen observed outbursts show evidence of a decreasing period over time, with a mean period of $P_0 = 114.2\pm0.4$ days and a period derivative of $\dot{P} = -0.0017\pm0.0003$. The most recent outburst in May 2020, which took place as predicted, exhibited spectroscopic changes during the rise and a had a UV bright, blackbody spectral energy distribution similar to tidal disruption events (TDEs). The X-ray flux decreased by a factor of 4 at the beginning of the outburst and then returned to its quiescent flux after $\sim8$ days. \textit{TESS} observed an outburst during Sectors 4-6, revealing a rise time of $5.60 \pm 0.05$ days in the optical and a decline that is best fit with an exponential model. We discuss several possible scenarios to explain ASASSN-14ko's periodic outbursts, but currently favor a repeated partial TDE. The next outbursts should peak in the optical on UT $2020\text{-}09\text{-}7.4 \pm 1.1$ and UT $2020\text{-}12\text{-}26.5 \pm 1.4$. 

\end{abstract}


\section{Introduction}

There are numerous physical processes that lead to variability in the nuclei of galaxies. Every massive galaxy likely houses a supermassive black hole (SMBH; \citealt{kormendy95}, \citealt{kormendy2013}), and the past several decades have been spent unraveling their accretion and variability processes (for reviews, see, e.g., \citealt{ulrich1997}, \citealt{ho08}, \citealt{heckman14}, \citealt{yuan14}, \citealt{padovani17},  \citealt{hickox18}, \citealt{komossa18}, \citealt{blanford19}). Without the ability to spatially resolve the immediate vicinity of the SMBH, other methods must be used to probe accretion physics.

Variable accretion in active galactic nuclei (AGN) is the primary driver of nuclear variability. Most quasars appear to vary in brightness stochastically with statistical properties that can be modeled relatively well by a Damped Random Walk (DRW; e.g., \citealt{kelly08}, \citealt{kozlowski10}, \citealt{macleod10}, \citealt{zu13}). Signatures that deviate from DRW behavior, namely periodic or semi-periodic features, have been suggested as possible indicators for a binary system at the galaxy's core (e.g., \citealt{komossa2006}). 

This has led to searches for periodic signals in AGN light curves to identify SMBH binaries. For example, \citet{graham2015crts} used the Catalina Real-time Transient Survey (CRTS) to search for sub-parsec SMBH binaries. They reported 111 candidates that showed evidence of periodicity associated with a Keplerian orbit. Another study by \citet{charisi2016} used the  Palomar Transient Factory to identify 33 candidates with evidence for periodic variability. Liu et al. (\citeyear{Liu2015,liu2016,Liu2019}) searched for periodicity in Pan-STARRS1's Medium Deep Survey, ultimately finding one candidate, PSO J185. Other candidates with quasi-periodic/periodic variability include NGC 4151 with an estimated period of $\sim$16 years (e.g., \citealt{oknyanskij1978}, \citealt{pacholczyk1983}, \citealt{guo2006}, \citealt{oknyanskij2007},  \citealt{bon2012}), and  PG 1302-102 with an estimated period of 1,884 days \citep{graham2015}. Simulations have shown that periodic variability is expected in the light curves of SMBH binaries at sub-parsec separations due to a variety of processes, including modulated mass accretion onto the binary (e.g., \citealt{dorazio13}, \citealt{gold14}, \citealt{farris14}) or relativistic Doppler boosting of the minidisks formed as a result of the binary interaction \citep{dorazio15}.

Aside from low-level variability, AGN can also show outbursts or flares in which the brightness of the AGN increases dramatically for a short period of time before returning to a level of relative quiescence. The best example of quasi-periodic optical flares is the 12-year outburst cycle of OJ 287. These were first reported by \citet{sillanpaa88}, who suggested that outbursts are due to perturbations of the primary black hole's accretion disk during pericenter passages of the secondary black hole on a 12 year orbital cycle. Relativistic effects like precession probably alter the orbital geometry so that the events are not strictly periodic (see, \citealt{valtonen06}, \citealt{Laine2020}). The most recent flare detected from OJ 287 brightened in the X-ray, UV, and optical and occurred between April-June 2020, which was consistent with the predictions of the binary black hole model \citep{komossa2020}. Another candidate is IC 3599, which has competing theories for the cause of its X-ray/optical flares. \citet{grupe15} proposed accretion disk instabilites as the cause of the recurring $\sim$9.5 year flares, while \citet{campana15} argued for partial tidal disruption events. 

Tidal disruption events (TDEs) also lead to variable SMBH activity. In a TDE, a star is torn apart when it passes within the tidal radius of the SMBH at the center of its host galaxy, as the tidal shear forces overwhelm the self-gravity of the star. The disrupted stellar material subsequently produces a luminous transient flare of electromagnetic radiation that we observe as a TDE. Total disruptions from parabolic orbits (\citealt{hills75}; \citealt{rees88}; \citealt{evans89}; \citealt{phinney89}) lead to half of the disrupted star's mass being ejected, while the other half is gravitationally bound and asymptotically returns to pericenter at a fallback rate proportional to $t^{-5/3}$. 

In the case of a partial TDE, the star survives the encounter with the SMBH and only a fraction of the stellar material is tidally stripped, leaving the stellar core intact. \citet{guillochon13} found that fallback rate for partial TDEs at late times becomes steeper than $t^{-5/3}$ because there is less debris with orbital binding energies close to zero if the stellar core survives the encounter. \citet{coughlin2019} found a fallback rate proportional to $t^{-9/4}$, which is effectively independent of the mass of the core that survives the passage close to the black hole. This fallback rate is supported by the hydrodynamical simulations of \citet{miles2020}. Hydrodynamical simulations also indicate that partial disruptions can repeat, causing episodic mass transfer from the star to the SMBH at every pericenter passage, resulting in a series of low-level flares that repeat on the orbital timescale \citep{macleod13}. Partial disruptions are most easily achieved for giant stars (e.g., \citealt{macleod13}, \citealt{guillochon13}) which might also be created as stellar merger products (e.g., \citealt{antonini2011}, \citealt{macleod12}).

Most theoretical predictions for TDEs predate any observations of the phenomenon. The first observational claims of TDEs were soft X-ray outbursts from otherwise quiescent galaxies (e.g., \citealt{bade96}, \citealt{komossagreiner1999}, \citealt{grupe99}, \citealt{greiner2000}, \citealt{gezari2003}, \citealt{komossa15}). Since then, TDE flares have been discovered at a range of wavelengths, including the hard X-ray (e.g., \citealt{bloom11}; \citealt{burrows11}; \citealt{cenko12}; \citealt{pasham15}), soft X-ray (e.g., \citealt{komossa99}; \citealt{donley02}; \citealt{maksym10}; \citealt{saxton12}), ultraviolet (e.g., \citealt{stern04}, \citealt{gezari06}, \citealt{gezari08}, \citealt{gezari09}), and optical (e.g., \citealt{vanvelzen11}, \citealt{gezari12}, \citealt{cenko12a}, \citealt{arcavi14}, \citealt{holoien14b}, \citealt{vinko15}, \citealt{holoien16b}, \citealt{holoien16}, \citealt{holoien2018}, \citealt{holoien19}, \citealt{holoien2019ps18kh}, \citealt{hinkle202019dj}, \citealt{holoien2020}, \citealt{vanvelzen2020}). Due to the intrinsic multi-wavelength properties of both TDEs and AGN, one problem is to identify characteristics that clearly distinguish between the two objects, notably using X-rays \citep{auchettl2018}. This is becoming more important with the discoveries of more ambiguous transients such as ASASSN-18el (\citealt{trakhtenbrot2019}, \citealt{ricci2020}) and ASASSN-18jd \citep{neustadt2020}.

Here we report the discovery and long-term observation of a series of periodic outbursts from ASASSN-14ko, which is associated with the AGN \gal{} (z=0.042489, \citealt{aguero96}). \gal{} was spectroscopically classified as a Type 2 Seyfert by \citet{lauberts82}.  In Section \ref{data} we discuss the discovery of ASASSN-14ko and the photometric and spectroscopic data used in this analysis. In Section \ref{bhmass} we discuss the host properties, and in Section \ref{analysis} we discuss the light curve and the periodic nature of the outbursts. The spectroscopic results are presented in Section \ref{spectraanalysis}, and we discuss several theoretical interpretations of these periodic flares in Section \ref{discussion}. For a flat $\Omega_m = 0.3$ universe, the luminosity distance is $\approx 188~\rm{Mpc}$ and the projected scale is $\approx 0.85 ~\rm{kpc}/\rm{arcsec}$. The Galactic extinction is $\text{A}_{\text{V}} = 0.118$ mag \citep{schlafly11}.

\section{Discovery and Observations} \label{data}

On 2014-11-14.28 UT, the All-Sky Automated Survey for Supernovae (ASAS-SN, \citealt{shappee14}, \citealt{kochanek17}) triggered on a nuclear transient associated with \gal{} at $V\sim17.0$ mag and reported it as ASASSN-14ko \citep{holoien14ATELc, holoien17}. We will refer to this object as ASASSN-14ko throughout this paper. As reported in \citet{holoien14ATELc}, a follow-up spectrum on 2014-11-16 using the Boller and Chivens (B\&C, \citealt{osip2004}) spectrograph on the du Pont 2.5m at Las Campanas Observatory revealed a strong blue continuum and the emission lines were consistent with a Type 2 Seyfert. At the time, the event was considered to be a Type IIn supernova with a blue continuum projected very close to the nucleus of a Type 2 Seyfert, but strong AGN activity was not ruled out as a possibility \citep{holoien14ATELc}.

UVOT and XRT observations by the Neil Gehrels Swift Observatory (\textit{Swift} hereafter, \citealt{gehrels04}) were taken on UT 2014-11-16, 2014-11-19, 2014-11-21, 2014-11-23, and 2014-11-27 (PI: Holoien, ToO ID: 33529). These observations showed that the central region of the galaxy had significantly brightened in the UV but were consistent with archival magnitudes in the optical. These \textit{Swift} data also revealed X-ray emission, with fluxes of (2.85$\pm$0.8)$\times$10$^{-13}$ ergs cm$^{-2}$ s$^{-1}$ and (3.1$\pm$0.7)$\times$10$^{-13}$ ergs cm$^{-2}$ s$^{-1}$ on UT 2014-11-16 and UT 2014-11-19, respectively. The X-ray spectrum was consistent with a highly absorbed AGN with a column density of $\sim$10$^{23}$ cm$^{-2}$ and a luminosity of L$_\mathrm{X}$ $\sim$ 3 $\times$10$^{42}$ ergs s$^{-1}$ \citep{holoien14ATELc}.

\begin{figure*}
    \centering
    \includegraphics[width=\linewidth]{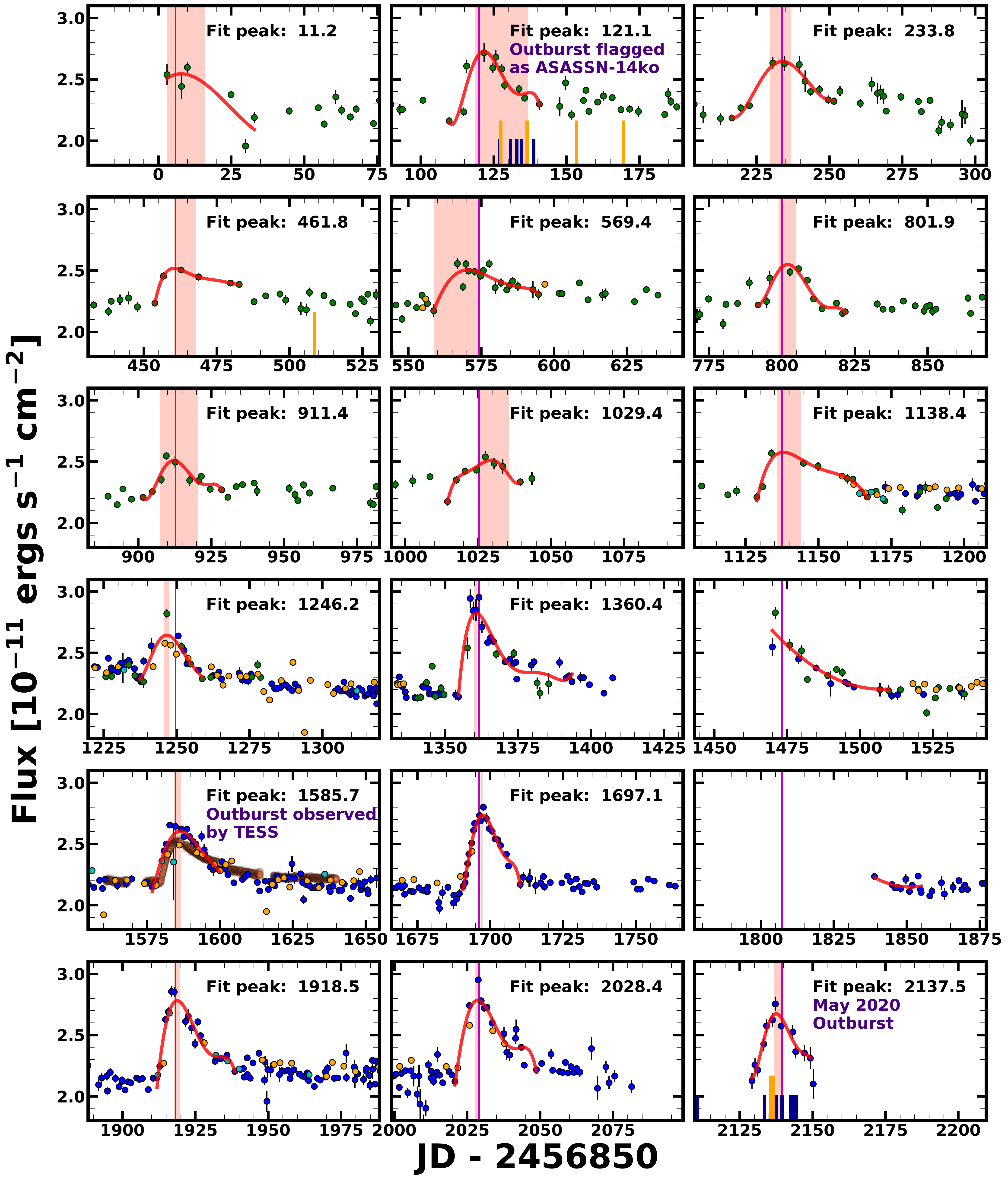}
    \caption{Light curves of ASASSN-14ko spanning 2014-2020 and demonstrating its periodic flaring behavior. The ASAS-SN $V$- and $g$-band data are shown in green and blue, respectively. The three \textit{TESS} sector observations are included in light red. ATLAS $o$-band data are shown in orange and ATLAS $c$-band data are shown in cyan. \textit{Swift} epochs are denoted by dark blue marks, and spectral epochs are shown by orange marks at the bottom of each panel. The fifth-order polynomial fits for each outburst are shown in red along with shaded red regions corresponding to estimates for each peak time and its uncertainty. The magenta vertical lines show the predicted peaks for the model with a period derivative described in Section \ref{pdot}. }
    \label{fig:ASASSN_LC}
\end{figure*}

As part of ongoing work to examine the long-term behavior of AGN observed by ASAS-SN, a full light curve of \gal{} was extracted in February 2020. Visual examination of the light curve revealed sixteen flares spaced out roughly evenly over six years, as shown in Figure \ref{fig:ASASSN_LC}. The seventeenth outburst in Figure \ref{fig:ASASSN_LC} was then predicted and observed. The original ASASSN-14ko trigger corresponds to the second outburst in the series. This initiated the further analysis and photometric and spectroscopic follow-up of ASASSN-14ko which we report here. All photometric data used in this analysis are presented in Table \ref{tab:all_phot}. 

\subsection{ASAS-SN Photometry}

ASAS-SN is a network of 20 robotic telescopes hosted by Las Cumbres Observatory Global Telescope (LCOGT, \citealp{brown13}) at five sites around the globe. Each telescope consists of four 14-cm aperture Nikon telephoto lenses with $8.\!\!''0$ pixels and a 4.5$^{\circ}$ $\times$ 4.5$^{\circ}$ field of view. ASAS-SN's primary objective is to discover supernovae by surveying the entire visible sky every night. The ASAS-SN data shown in Figure \ref{fig:ASASSN_LC} includes both $V$-band and $g$-band observations. In 2018, the first two ASAS-SN mounts transitioned from $V$-band to $g$-band to match the three ASAS-SN telescopes deployed in 2017-2018. 

The data were reduced using a fully-automated pipeline based on the ISIS image subtraction package \citep{alard98, alard00}. Each photometric epoch (usually) combines three dithered 90-second image exposures with a 4.47$\times$4.47 degree field-of-view that is subtracted from a reference image.  We then used the IRAF package \texttt{apphot} to perform aperture photometry with a 2-pixel, or approximately $16.\!\!''0$, radius aperture on each subtracted image, generating a differential light curve. The photometry was calibrated using the AAVSO Photometric All-Sky Survey \citep{henden15}. All low-quality ASAS-SN images of \gal{} were inspected by-eye, and images  with clouds or other systematic problems were removed.

\begin{table}
\centering
\caption{Photometry of ASASSN-14ko used in this analysis. Only the first observation in each band is shown here to demonstrate its form and content. Table to be published in its entirety in machine-readable form in the online journal.   }
\begin{tabular*}{\columnwidth}{l l l l}
\toprule

JD  & Band & \pbox{20mm}{Flux ($10^{-11}$\\ $\mathrm{ergs}\ \mathrm{s}^{-1}\ \mathrm{cm}^{-2}$)}  & \pbox{25mm}{ Flux Error ($10^{-11}$ \\ $\mathrm{ergs} \ \mathrm{s}^{-1} \mathrm{cm}^{-2}$)} \\

\hline

2458957.425 & X-ray & 0.18 & 0.03 \\
2458957.428 & $UVW2$ & 1.74 & 0.06 \\
2458957.431 & $UVM2$ & 1.30 & 0.04 \\
2458957.425 & $UVW1$ & 1.36 & 0.05 \\
2458957.427 & $U$ & 1.26 & 0.04 \\
2458952.424 & $B$ & 1.88 & 0.03 \\
2458022.900 & $g$ & 2.29 & 0.05 \\ 
2456790.495 & $V$ & 2.24 & 0.07 \\ 
2457453.755 & $c$ & 1.30 & 0.01 \\
2457404.879 & $o$ & 2.20 & 0.01 \\
2458411.127 & $I_{TESS}$ & 0.10 & 0.01 \\
2458958.385 & $r$ & 2.65 & 0.02 \\

\hline 

\end{tabular*}
\label{tab:all_phot}
\end{table}

\subsection{Swift UVOT Photometry}

Following the original discovery, we requested \textit{Swift} UltraViolet/Optical Telescope (UVOT, \citealp{roming05}) ToO observations (ToO ID: 33529). Then, after we discovered its periodic nature, we again requested  \textit{Swift} data (ToO IDs: 13836, 13979, 14005) to monitor ASASSN-14ko during  quiescence and then during the outburst predicted for UT 2020-05-18.5 (see below). Data were obtained in  six  filters  \citep{poole08}: $V$ (5468 \AA), $B$ (4392 \AA), $U$ (3465 \AA), $UVW1$  (2600 \AA), $UVM2$  (2246 \AA),  and $UVW2$ (1928 \AA). We used the HEAsoft (\hspace{-1mm}\citealt{heasarc2014}) software task \textit{uvotsource} to extract the source counts using a $16.\!\!''0$ radius aperture and used a sky region of $\sim$ $40.\!\!''0$ radius to estimate and subtract the sky background. This aperture size was chosen to match the ASAS-SN photometry. All fluxes were aperture corrected and converted into magnitudes and fluxes using the most recent UVOT calibration (\citealt{poole08}, \citealt{breeveld10}). The UVOT transient, host, and calibration star magnitudes were corrected for Galactic extinction. In order to measure only the transient flux in each epoch, the quiescent host fluxes were subtracted in the same aperture. The host magnitudes are given in Table \ref{tab:hostmags}. We converted the \textit{Swift} UVOT $B$ and $V$ magnitudes to Johnson $B$ and $V$ magnitudes using the standard color corrections\footnote{\url{https://heasarc.gsfc.nasa.gov/docs/heasarc/caldb/swift/docs/uvot/uvot_ caldb_coltrans_02b.pdf}}. 

There are three additional observations of this galaxy in the \textit{Swift} data archive under the identification SWIFT J$0525.3-4600$ (Obs ID: 37354, PI: Markwardt). These observations were obtained to investigate the \textit{Swift} Burst Alert Telescope (BAT) source reported by \citet{baumgartner13} in their 70 month catalog and identified with the nearby blazar $PKS 0524-460$. However, \citet{lansbury2017} concluded that \gal{} was in fact the BAT source rather than PKS $0524-460$. The three \textit{Swift} epochs are consistent with the quiescent magnitudes we measure from the later data and there were too few observations to usefully add to the constraints on the times of the outbursts.

\begin{table}
\centering
    \caption{Host magnitudes of \gal{} measured using a $16.\!\!''0$ aperture. The $UVW2$-, $UVM2$-, $UVW1$- and $U$-band magnitudes were determined from Swift data taken in April 2020 during pre-outburst quiescence. Johnson-Cousins $BVR$ and SDSS $gri$ magnitudes were determined from LCOGT, Swift $B$ and $V$, and amateur astronomer data also taken in April 2020. The magnitudes were combined by averaging the data weighted by the inverse squares of the uncertainties. All magnitudes are in the AB system.}
    \begin{tabular*}{0.63\columnwidth}{l l l}
\toprule

Filter & Magnitude & Uncertainty \\

\hline

$UVW2$ & 16.26 & 0.02 \\
$UVM2$ & 16.37 & 0.02 \\
$UVW1$ & 16.17 & 0.02 \\
$U$ & 15.95 & 0.02 \\
$B$ & 15.26 & 0.01 \\
$g$ & 15.00 & 0.01 \\
$V$ & 14.77 & 0.01 \\
$r$ & 14.46 & 0.01 \\
$R$ & 14.48 & 0.01 \\
$i$ & 14.68 & 0.01 \\
$J$ & 13.45 & 0.04 \\
$H$ & 12.65 & 0.05 \\ 
$K_s$ & 11.76 & 0.04 \\
$W1$ & 12.78 & 0.02 \\
$W2$ & 12.24 & 0.02 \\
$W3$ & 10.40 & 0.02 \\
$W4$ & 9.11 & 0.02 \\
\hline 

\end{tabular*}
    \label{tab:hostmags}
\end{table}

\subsection{ATLAS Photometry} 
The ATLAS survey \citep{tonry18} consists of two 0.5m f/2 Wright Schmidt telescopes on Haleakal\={a} and at the Mauna Loa Observatory. Designed primarily for detecting hazardous asteroids, the telescopes obtain four 30-second exposures of 200-250 fields per night. This corresponds to roughly a quarter of the visible sky. ATLAS uses two broad-band filters, the `cyan' (\textit{c}) filter covering 420 - 650 nm and the `orange' (\textit{o}) filter covering 560 - 820 nm \citep{tonry18}. 

The ATLAS pipeline performs flat-field corrections for each image as well as astrometric and photometric calibrations. Reference images of the host galaxy were created by stacking multiple images taken under ideal conditions and this reference was then subtracted from each ATLAS epoch to isolate the flux from the transient. We performed forced photometry on the subtracted ATLAS images of ASASSN-14ko as described in \citet{tonry18}. ATLAS images taken on the same night were stacked. The resulting ATLAS \textit{o}- and \textit{c}- band photometry and 3-sigma limits are included also in Figure \ref{fig:ASASSN_LC}.

\subsection{\textit{TESS} Photometry}

ASASSN-14ko was observed by the Transiting Exoplanet Survey Satellite (\textit{TESS}, \citealt{ricker2014}) during Sectors 4-6, which occurred between 2018-10-18 and 2019-01-07. Similar to the process applied to the ASAS-SN data, we used the ISIS package \citep{alard98,alard00} to perform image subtraction on the 30-minute cadence \textit{TESS} full frame images (FFIs) to obtain high fidelity light curves of this galaxy. This process is fully described in \citet{vallely2019}. 

We construct independent reference images for each sector as opposed to utilizing a single reference image over all sectors to avoid introducing problems created by the field rotations between sectors. The reference images were built using the first 100 good-quality FFIs for each sector. FFIs were considered poor-quality if the sky background levels or PSF widths were above average for the sector. FFIs were also excluded from our analysis if they had data quality flags, or were taken when the spacecraft's pointing was compromised due to instrument anomalies, or when scattered light affected the images. 

The measured fluxes were converted into \textit{TESS}-band magnitudes using an instrumental zero point of 20.44 electrons per second from the \textit{TESS} Instrument Handbook \citep{TESSHandbook}. \textit{TESS} observes in a single broad-band filter, spanning roughly 6000--10000\,\AA{} with an effective wavelength of $\sim$7500\,\AA{}, and \textit{TESS} magnitudes are calibrated to the Vega system \citep{sullivan15}. The \textit{TESS} light curve is also shown in Figure \ref{fig:ASASSN_LC}.

\subsection{Las Cumbres Observatory Global Telescope Photometry}

We obtained photometric observations from Las Cumbres Observatory Global Telescope (LCOGT, \citealt{brown13}). The $B$-, $V$-, $g'$-, $r'$-, and $i'$-band observations were taken using the 1-meter telescope at Siding Spring Observatory in New South Wales, Australia. The LCOGT photometric observations began on 2020-04-13 in quiescence and continued through the midpoint of the May 2020 outburst when telescope horizon observing limits prevented further observations. Aperture magnitudes were obtained using a $16.\!\!''0$ radius aperture using the IRAF \texttt{apphot} package using an annulus to estimate and subtract background counts. We used stars with APASS DR 10 magnitudes to calibrate the data. Similar to the process for the UVOT observations, the aperture magnitudes were corrected for Galactic extinction. The host galaxy flux was measured using the April 2020 quiescent data and subtracted to isolate the flux from the transient.

\subsection{Amateur Astronomer Photometry}

Amateur astronomers at four different observatories observed ASASSN-14ko starting shortly after it was discovered to be periodically flaring. Data were taken at Moondyne Observatory, east of Perth, Australia, using a 0.4-m telescope with AOX adaptive optics between April 28 and June 8, 2020 and on a daily cadence between May 10 and May 28, 2020. $B_C$-, $V$-, $G_S$-, $R_S$-, and $I_C$-band images were obtained with guided 120 second exposures. The data were reduced and calibrated with standard procedures, and then stacked with 3 and 5 image sets aligned using background stars. Data were also collected using a 41-cm telescope at Savannah Skies Observatory from Queensland, Australia. The bias and dark subtracted data were taken in the $B$-, $V$-, $R_C$-, and $I_C$-bands using 180 second exposures. ASASSN-14ko was observed from Bronberg Observatory in South Africa using 14- and 12-inch telescopes in the $R$ band with 15 second exposures. The images were calibrated and then sets of eight images were stacked per night. Finally, ASASSN-14ko was observed from Dogsheaven Observatory in Brazil using the $B$, $V$, $R_C$, and $I_C$ filters. The images were observed with 120 second exposures using a 14-inch telescope and calibrated. 

All these images were then astrometrically calibrated and aperture magnitudes were measured using the IRAF  \texttt{apphot} package and a $16.\!\!''0$ aperture following the same procedures described for the \textit{Swift} UVOT and LCOGT data. 

\subsection{Spectroscopic Observations} 

The first available observation of \gal{} was obtained by \citet{kewley01} on 1996-02-19. At that time, \gal{} was classified as a Seyfert 2 galaxy. When ASASSN-14ko was first discovered, we took a follow-up spectrum using the B\&C spectrograph on the du Pont 2.5-m at Las Campanas Observatory on 2014-11-16. Other spectra were taken by PESSTO \citep{smartt15} as part of transient follow-up using the ESO-NTT/EFOSC2-NTT on 2014-11-25, 2014-12-12, 2014-12-28, 2015-01-26, and 2015-01-27. These spectra are available at WISEREP \citep{yaron2012} and in the ESO archive\footnote{http://archive.eso.org}. Both spectra taken in November 2014 showed noticeably broadened Balmer emission-lines compared to the archival spectrum from 1996.  

We also obtained seven spectra with the LCOGT FLOYDS spectrograph \citep{sand2014} at the robotic 2-m Faulkes Telescope South located at Siding Spring Observatory \citep{brown13}. These observations were taken on 2020-04-12, 2020-04-15, 2020-04-24, 2020-04-25, 2020-04-27, 2020-05-16, and 2020-05-17 in order to observe any changes in spectral features prior to and during the most recent outburst. All spectra were reduced following standard reduction procedures using IRAF. All observations were taken with an exposure time of 600 seconds and span 4,300 to 10,000\AA{}.

We used the analysis tool \texttt{mapspec}\footnote{\href{https://github.com/mmfausnaugh/mapspec}{https://github.com/mmfausnaugh/mapspec}} (MCMC Algorithm for Parameters of Spectra; \citealt{fausnaugh16}) to calibrate the longslit spectra onto the same absolute flux scale using the [OIII]$\lambda$ 5007 flux to scale the spectra. We assume that the [OIII] narrow-line flux is constant because it originates in an extended region too large to vary on these short time scales. \texttt{mapspec} uses MCMC methods to adjust the flux, wavelength shift, and resolution of each individual spectrum to match that of the reference spectrum. The reference spectrum was defined by an average of the spectra sample. 

\subsection{X-ray data}

In addition to the \textit{Swift} UVOT observations, we also obtained simultaneous \textit{Swift} X-Ray Telescope (XRT, \citealt{burrows2005}) photon-counting observations of ASASSN-14ko. All observations were reprocessed from level one XRT data using the \textit{Swift} \textsc{xrtpipeline} version 0.13.2, producing cleaned event files and exposure maps. Standard filter and screening criteria\footnote{\url{http://swift.gsfc.nasa.gov/analysis/xrt_swguide_v1_2.pdf}} were used, as well as the most up-to date calibration files.

To extract background-subtracted count rates, we used a source region with a radius of 50$''$ centered on the optical position of ASASSN-14ko. To define the background, we used a $150.\!\!''0$ radius source free region centered at ($\alpha$,$\delta$)=($05^{h}25^{m}18.08^{s},-46^{\circ}00''21.0'$). All count rates are aperture corrected.

To improve the signal to noise, we merged the most recent \textit{Swift} XRT observations (ObsIDs: 14005, 13979, 13836) using the HEASOFT tool \textsc{xselect}. From this merged observation, we extracted spectra using the task \textsc{xrtproducts} version 0.4.2 and the same extraction regions. Ancillary response files were obtained by merging the individual exposure maps using \textsc{XIMAGE} version 4.5.1 and the task \textsc{xrtmkarf}. We used the ready-made response matrix files that are available with the \textit{Swift} calibration files. This merged \textit{Swift} spectrum was grouped to have a minimum of 10 counts per energy bin using the \textsc{FTOOLS} command \textsc{grppha}.

On 2015 August 19, \gal{} was observed using the MOS and PN detectors onboard \textit{XMM-Newton} (ObsID: 0762920501, PI: Koss) as part of a program to study heavily obscured AGN. Both of the detectors were operated in full-frame mode using a thin filter. We reduced the data using \textit{XMM-Newton} science system version 15.0.02 and the most up to date calibration files. Periods of high background/proton flares that could affect the quality of the data were identified by generating a count rate histogram of events between 10 and 12 keV. The observations were only marginally affected by background flares, leading to effective exposures of 25.6\,ks and 24.0\,ks for the MOS and PN  detectors, respectively. For our analysis, we used standard event screening and flags for both the MOS and PN detectors\footnote{\url{https://xmm-tools.cosmos.esa.int/external/xmm_user_support/documentation/sas_usg/USG.pdf}}. All files were corrected for vignetting using \textsc{EVIGWEIGHT}. Spectra were extracted from both detectors using the SAS task EVSELECT and the cleaned event files. We used the same source region used to analyze the\textit{ Swift} observations, and the spectra were grouped using \textsc{grppha} to have a minimum of 20 counts per energy bin.

\gal{} was also observed using the Nuclear Spectroscopic Telescope Array (\textit{NuSTAR}) on 2015 August 21  (ObsID: 60101014002, PI: Koss) as part of the same program to observe heavily obscured AGN. We reduced the data using the {\it NuSTAR} Data Analysis Software (NuSTARDAS) Version 1.8.0 and {\it NuSTAR} CALDB Version 20170817. We performed the standard pipeline data processing with {\it nupipeline}, with the \texttt{saamode=STRICT} to identify the South Atlantic Anomaly passages. Using the {\it nuproducts} \ FTOOL, we extracted source spectra from a 100\arcsec-radius region and produced ancillary response files and redistribution matrix files for both the A and B modules. We extracted background spectra from annular regions centered on the source, and we followed the procedure outlined by \cite{wik14} to estimate the backgrounds and subtract them from the source spectra using the {\it nuskybgd} routines\footnote{https://github.com/NuSTAR/nuskybgd}.

To analyze the merged \textit{Swift},  \textit{XMM-Newton} and \textit{NuSTAR} spectra we used the X-ray spectral fitting package (XSPEC) version 12.10.1f and $\chi^2$ statistics. Finally, to further constrain the X-ray emission, we also analysed the available \textit{XMM-Newton} slew observations of the region. \textit{XMM-Newton} slew observations take advantage of the fast read out of the PN detector and its ability to observe the sky without reduction of image quality as \textit{XMM-Newton} maneuvers between pointed observations. Slew observation can detect X-ray emission down to a 0.2-10.0 keV flux limit of $\sim10^{-12}$ ergs cm$^{-2}$ s$^{-1}$ \citep{saxton08}. We found nine slew observations overlapping the source and analyzed them using the SAS tool \textsc{eslewchain}\footnote{\url{https://www.cosmos.esa.int/web/xmm-newton/sas-thread-epic-slew-processing}}. To extract count rates, we use the same source and background regions used for the pointed \textit{XMM-Newton} observation. Due to the low exposure times of each observation no spectra could be extracted.

\section{Host Properties and SMBH Mass} \label{bhmass}
We fit the quiescent host photometry given in Table \ref{tab:hostmags} using \texttt{AGNfitter} \citep[][]{calistrorivera16}. In addition to our photometry from Section \ref{data}, we also include the $J$, $H$, and $K_s$ fluxes from the 2MASS All-Sky Point Source Catalog \citep{skrutskie06}, the $W1$-$W4$ fluxes from the AllWISE Source Catalog \citep{wright2010}, and the 12, 25, 60, and 100$\mu m$ fluxes from the IRAS Faint Source Catalog \citep{moshir1990}. \texttt{AGNfitter} uses MCMC methods to model the combined contribution of the AGN accretion disk, dusty torus, stellar population, and cold dust. Based on our fit, \gal{} has a stellar mass of M$_* = (3.68^{+0.06}_{-0.05}) \times 10^{10}$ M$_{\odot}$, a stellar population age of $0.59 \pm 0.10$ Gyr, and a star formation rate of SFR = $24.4 \pm 0.3 $ M$_{\odot}$ yr$^{-1}$.

We used several approaches to estimate the SMBH mass. First, we used the 2014-11-25 PESSTO spectrum to estimate a broad-line region radius (R$_{BLR}$) using the scaling relation of \citet{kaspi05}. We then fit the broad component of H$\beta$ with a Gaussian to estimate the characteristic velocity of the BLR. Assuming a Keplerian orbit and using a geometry factor of 4.3 \citep{bentz15}, this yields a SMBH mass $\log_{10}(M_{BH}) = 8.18^{+0.05}_{-0.07}$ M$_{\odot}$. Second, we applied the scaling relation between bulge near-infrared (NIR) luminosity and SMBH mass of \citet{marconi03}. Assuming the AGN contributes 33\% of the total flux in the NIR (the median value in \citealp{burtscher15}), we used archival 2MASS photometry to estimate a SMBH mass of $\log_{10}(M_{BH}) = 7.89^{+0.15}_{-0.23}$ M$_{\odot}$, $\log_{10}(M_{BH}) = 7.65^{+0.18}_{-0.29}$ M$_{\odot}$, and $\log_{10}(M_{BH}) = 7.45^{+0.19}_{-0.37}$ M$_{\odot}$ for the $J$, $H$, and $K_s$ bands respectively. Finally, we used the host stellar mass from our SED fits to estimate a bulge mass following \citet{mendel14}. Then, from the M$_B$ - M$_{BH}$ relation of \citet{mcconnell13}, we estimated a black hole mass of $\log_{10}(M_{BH}) = 7.85^{+0.18}_{-0.34}$ M$_{\odot}$. A weighted average of these measurements with an uncertainty which incorporates the range of the estimates yields $\log_{10}(M_{BH}) = 7.86^{+0.31}_{-0.41}$ M$_{\odot}$. We assume this value throughout this manuscript. The corresponding Eddington luminosity for a black hole of this mass is $\log_{10}(L_{Edd}) = 45.95^{+0.32}_{-0.41}$ ergs s$^{-1}$.

\begin{table}[t]
\centering
\caption{ Observed time and maximum light of each outburst observed by ASAS-SN in the V- and g-band.  Shown in blue are the predicted times of the next two outburst peaks in the optical, based on our standard model with a period derivative.  }
\begin{tabular*}{0.9\columnwidth}{l l l}
\toprule

JD & Flux at Peak (mJy) & O-C (days) \\

\hline

2456861.2$^{+4.9}_{-8.3}$&4.24$^{+1.86}_{-0.24}$&$-$16.8$^{+4.9}_{-11.7}$\\  
2456971.1$^{+15.6}_{-2.6}$&4.53$^{+3.49}_{-0.13}$&$-$18.1$^{+15.6}_{-10.0}$\\  
2457083.8$^{+3.0}_{-4.2}$&4.26$^{+1.42}_{-0.06}$&$-$16.6$^{+3.0}_{-9.8}$\\  
2457311.8$^{+6.1}_{-1.5}$&4.00$^{+0.25}_{-0.01}$&$-$11.1$^{+6.1}_{-1.5}$\\  
2457419.4$^{+4.7}_{-10.7}$&4.07$^{+7.09}_{-0.07}$&$-$14.7$^{+4.6}_{-16.0}$\\  
2457651.9$^{+3.0}_{-3.0}$&4.14$^{+1.18}_{-0.14}$&$-$4.7$^{+3.1}_{-3.1}$\\  
2457761.4$^{+9.0}_{-3.8}$&4.07$^{+1.37}_{-0.07}$&$-$6.4$^{+9.1}_{-3.8}$\\  
2457879.4$^{+6.2}_{-4.1}$&4.06$^{+1.32}_{-0.16}$&$+$0.4$^{+6.2}_{-4.2}$\\  
2457988.4$^{+5.7}_{-2.5}$&4.19$^{+0.55}_{-0.19}$&$-$1.9$^{+5.7}_{-2.5}$\\  
2458096.2$^{+1.3}_{-0.5}$&4.33$^{+0.16}_{-0.13}$&$-$5.3$^{+1.4}_{-0.7}$\\  
2458210.4$^{+0.6}_{-0.6}$&4.40$^{+0.13}_{-0.10}$&$-$2.3$^{+0.8}_{-0.8}$\\  
2458435.7$^{+1.0}_{-1.4}$&4.08$^{+0.02}_{-0.08}$&$+$0.5$^{+1.1}_{-1.5}$\\  
2458547.1$^{+0.3}_{-0.2}$&4.24$^{+0.07}_{-0.04}$&$+$0.7$^{+0.6}_{-0.6}$\\  
2458768.5$^{+1.2}_{-0.8}$&4.33$^{+0.12}_{-0.13}$&$-$0.3$^{+1.3}_{-1.0}$\\  
2458878.4$^{+0.3}_{-0.6}$&4.39$^{+0.11}_{-0.09}$&$-$1.7$^{+0.6}_{-0.8}$\\  
2458987.5$^{+2.4}_{-0.8}$&4.17$^{+0.22}_{-0.09}$&$-$3.8$^{+2.4}_{-0.8}$\\  
{\color{blue}2459099.9$^{+1.1}_{-1.1}$}&{\color{blue}{...}}&{\color{blue}{...}}\\
{\color{blue}2459210.0$^{+1.4}_{-1.4}$}&{\color{blue}{...}}&{\color{blue}{...}}\\

\hline 

\end{tabular*}
\label{tab:peak_table}
\end{table}

\section{Light Curve Analysis} \label{analysis}

\subsection{Periodic Outbursts in the Light Curve} \label{pdot}

\begin{figure*}[t]
    \centering
    \includegraphics[width=\textwidth]{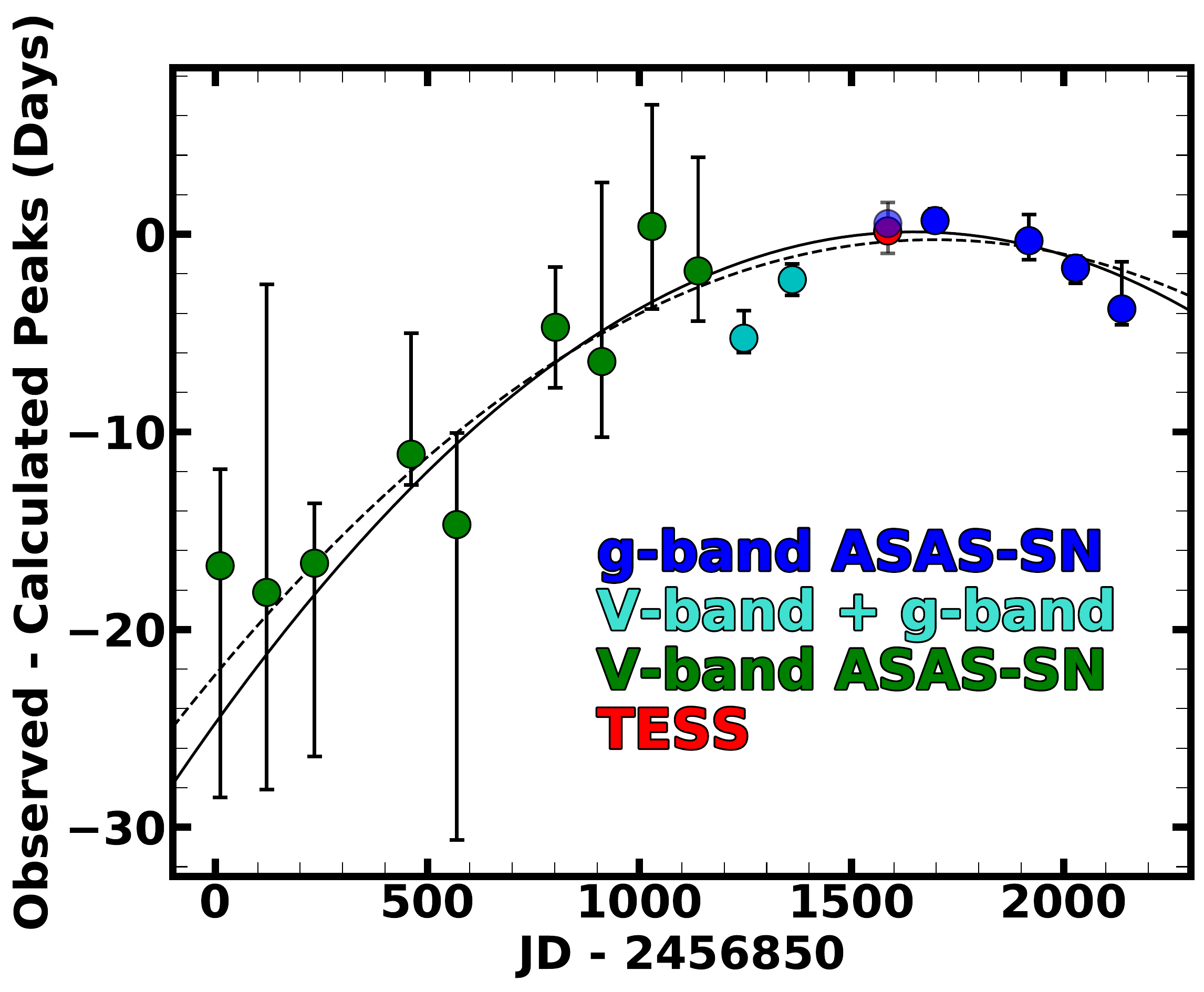}
    \caption{O-C plot comparing the observed peak time for each outburst with the estimated peak if we assume a constant period. The parabolas show the predictions from the models including a $\dot{P}$ either with the assumed (solid) or re-scaled (dashed) uncertainties for the times of peak. }
    \label{fig:o-c_plot}
\end{figure*}

We individually fit each outburst with a fifth-order polynomial to determine the timings of the peaks as shown in red in Figure \ref{fig:ASASSN_LC}. We then measured the errors on the peak times by bootstrap resampling the light curves. These errors are shown by the shaded red regions in Figure \ref{fig:ASASSN_LC}. The times and fluxes for each peak in the $V$- and $g$-band ASAS-SN light curve are given in Table \ref{tab:peak_table}.

Figure \ref{fig:ASASSN_LC} visually demonstrates that the outburst peaks recur at consistent intervals. Initially, we analyzed the light curves using the box least squares (BLS) periodogram method \citep{kovacs2002} to obtain a period of 111.82 $\pm$ 1.9 days from the first sixteen observed peaks. We used the Half-Width-at-Half-Maximum (HWHM) as an estimate for the period uncertainty \citep{mighell13}. This period estimate was used to predict and plan for the 2020-05-18.5 outburst. 

\begin{figure*}[t]
    \centering
    \includegraphics[width=\textwidth]{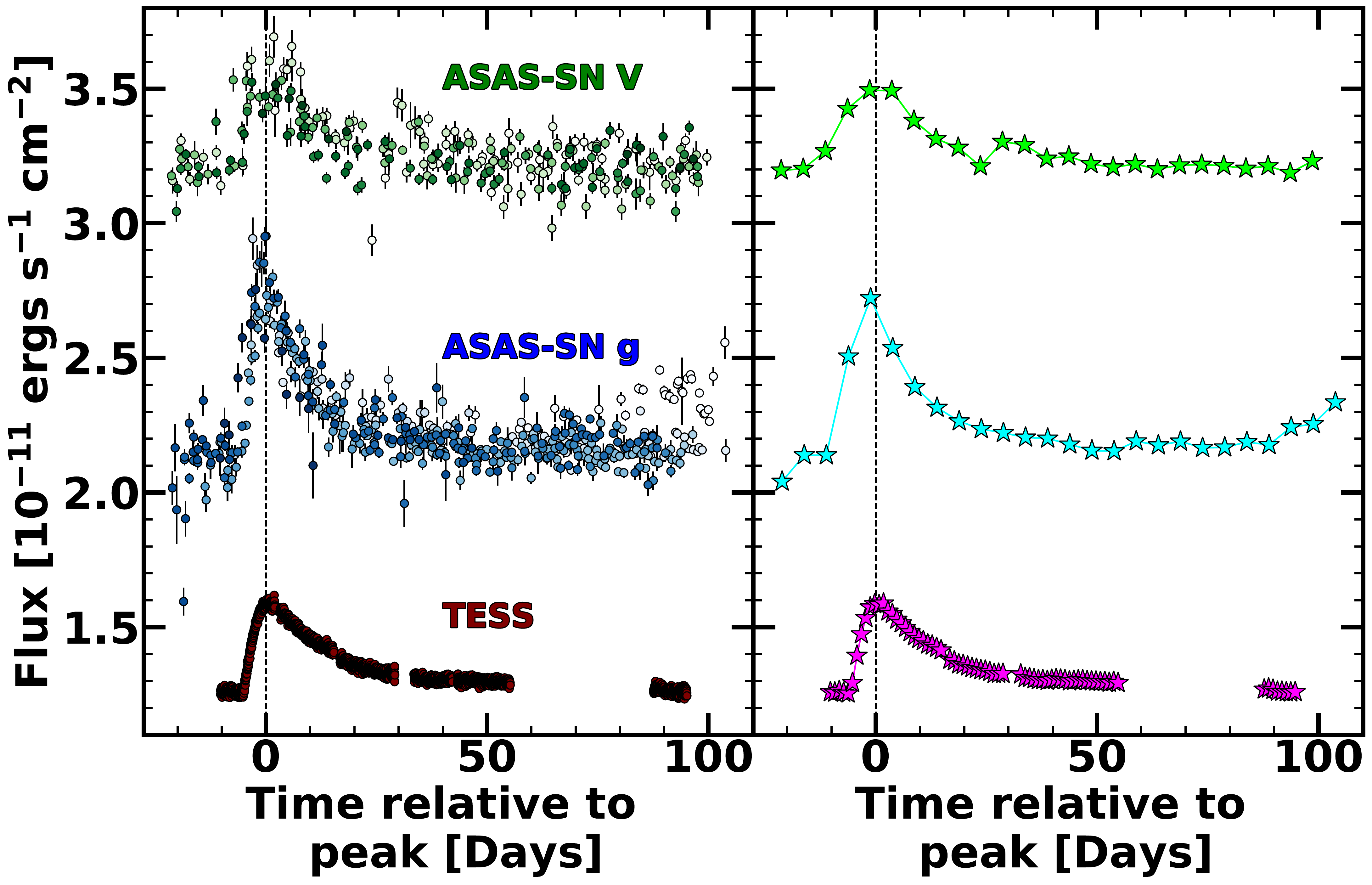}
    \caption{The stacked ASAS-SN $V$-band (green), ASAS-SN $g$-band (blue) and \textit{TESS} (maroon) light curves as a function of phase. The light curves are offset for clarity and the data of each phase are given a different color shade. The right panel shows the phase-stacked light curves binned at 5 day intervals for ASAS-SN and binned at 1 day intervals for \textit{TESS}. The outburst peaks align across these filters and the shapes are similar.}
    \label{fig:phasefold}
\end{figure*}

Using this mean period leads to significant residuals between the observed and calculated peak times, as shown in the Observed - Calculated (O-C) diagram of Figure \ref{fig:o-c_plot}. Such a periodic trend is indicative of a period derivative, so for the present analysis we fit the time of the transient peak arrivals as  

\begin{equation}
    t = t_0 + nP_0 + \frac{1}{2}n^2 P_0 \dot{P} + \frac{1}{6} n^3 P_0 \dot{P}^2,
\end{equation}
where $t_0$ is a reference time, $P_0$ is the ``mean" period, $\dot{P}$ is the period derivative, and $n$ is the peak number starting from the first peak set as n=0. We use MCMC estimates of the 1-$\sigma$ parameter uncertainties. When fitting without a  $\dot{P}$ initially, we found the best-fit parameter $P_0$ = 111.20 $\pm$ 0.06 days but the fit returns a reduced $\chi^2$ of 9.75 for 14 degrees of freedom. The fit noticeably improved when including a $\dot{P}$ with $t_0$ = 3.4$^{+2.2}_{-2.3}$ days, $P_0$ = 114.6 $\pm$ 0.3 days, $\dot{P}$ = $-$0.0021 $\pm$ 0.0002 and a reduced $\chi^2$ of 2.07. This fit is shown in Figure \ref{fig:o-c_plot} as the solid line. If we expand the time of peak uncertainties in quadrature by 0.63 days to have a $\chi^2$ per degree of freedom of unity, we find best-fit parameters $t_0$ = 5.9 $\pm$ 2.7 days, $P_0$ = 114.2 $\pm$ 0.4 days, and $\dot{P}$ = $-$0.0017 $\pm$ 0.0003. This fit is shown by Figure \ref{fig:o-c_plot} as the dashed line and we use it as our standard model. For this model, the next two outbursts are predicted to peak in the optical on MJD 59099.9 $\pm$ 1.1 and MJD 59210.0 $\pm$ 1.4, which correspond to 2020 September 7.4 $\pm$ 1.1 and 2020 December 26.5 $\pm$ 1.4.

The $V$- and $g$-band ASAS-SN light curves for ASASSN-14ko stacked using the phasing of this model are shown in Figure \ref{fig:phasefold} along with the \textit{TESS} light curve. Including the $\dot{P}$ significantly reduces the scatter between the stacked light curves. We also binned these stacked light curves. As apparent in the right panel of Figure \ref{fig:phasefold}, the peak times are closely aligned across these three filters and the light curve morphologies are similar between peaks and these filters. The outbursts are characterized by a fast rise to peak followed by a shallower decline.

The Catalina Real-Time Transient Survey (CRTS; \citealt{drake09}) data contain observations spanning nine years prior to the start of the ASAS-SN $V$-band light curve. However, the CRTS data quality made it difficult to identify prior outburst peaks relative to the quiescent baseline. The CRTS photometry includes flux from the entire host galaxy since it uses Source Extractor \citep{bertin1996} rather than image subtraction. Since the host galaxy is bright and spatially large, it is difficult to recognize prior outbursts. We also found that data from the All-Sky Automated Survey (ASAS; \citealt{pojmanski1997}) were not useful for identifying earlier outbursts and we could find no other earlier data.

\subsection{\textit{TESS} Light Curve Analysis} \label{tess_decline_rise}

\begin{figure*}[htp]
    \centering
    \includegraphics[width=0.95\linewidth]{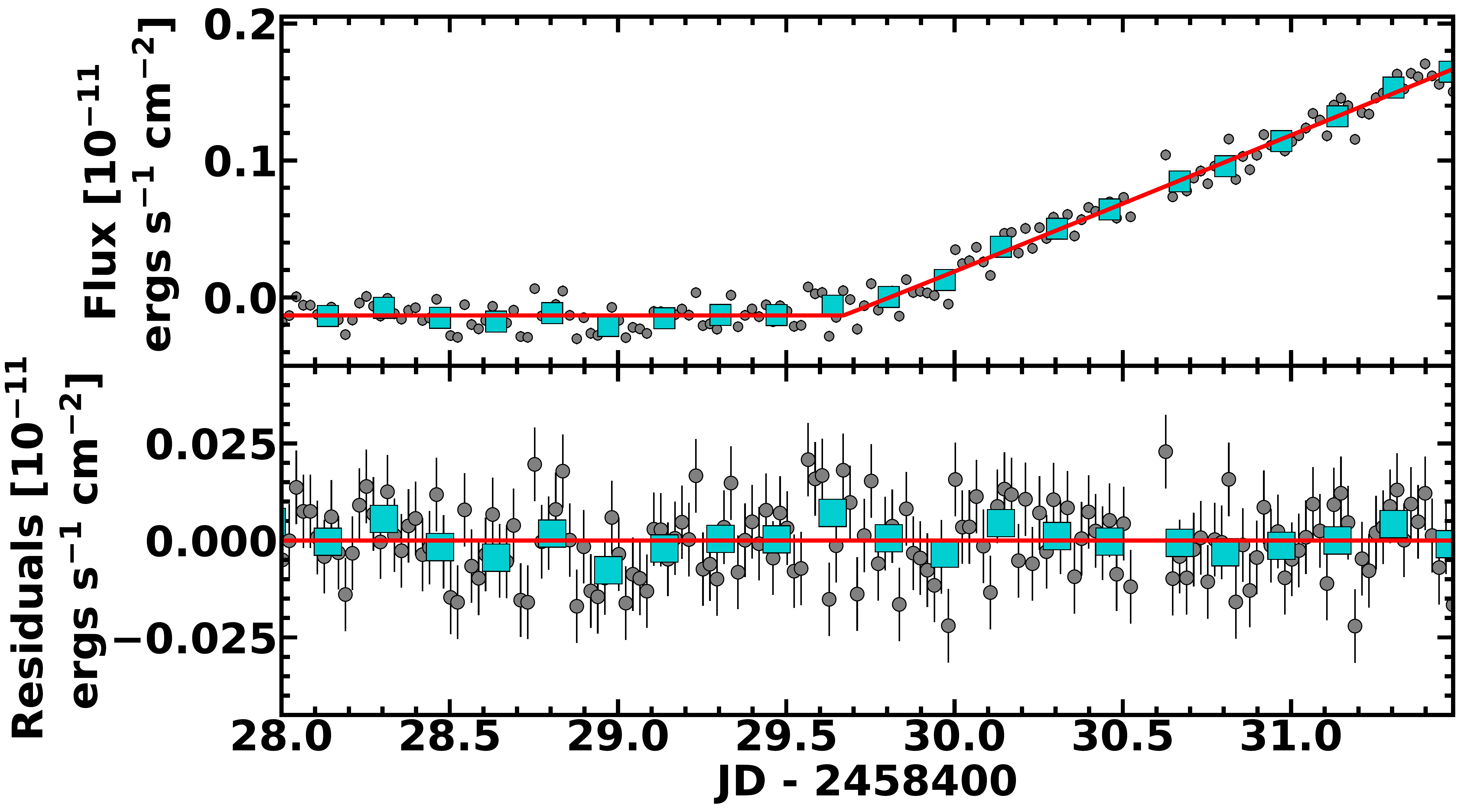}
    \caption{The rising phase of the \textit{TESS} image subtraction light curve. The best-fit power-law model for the rise until JD 2458431.5 is shown in red. The bottom panel shows the flux residuals. The \textit{TESS} data binned by four hours are shown by turquoise squares.}
    \label{fig:tess_rise_0.4mjy}

    \centering
    \includegraphics[width=0.95\linewidth]{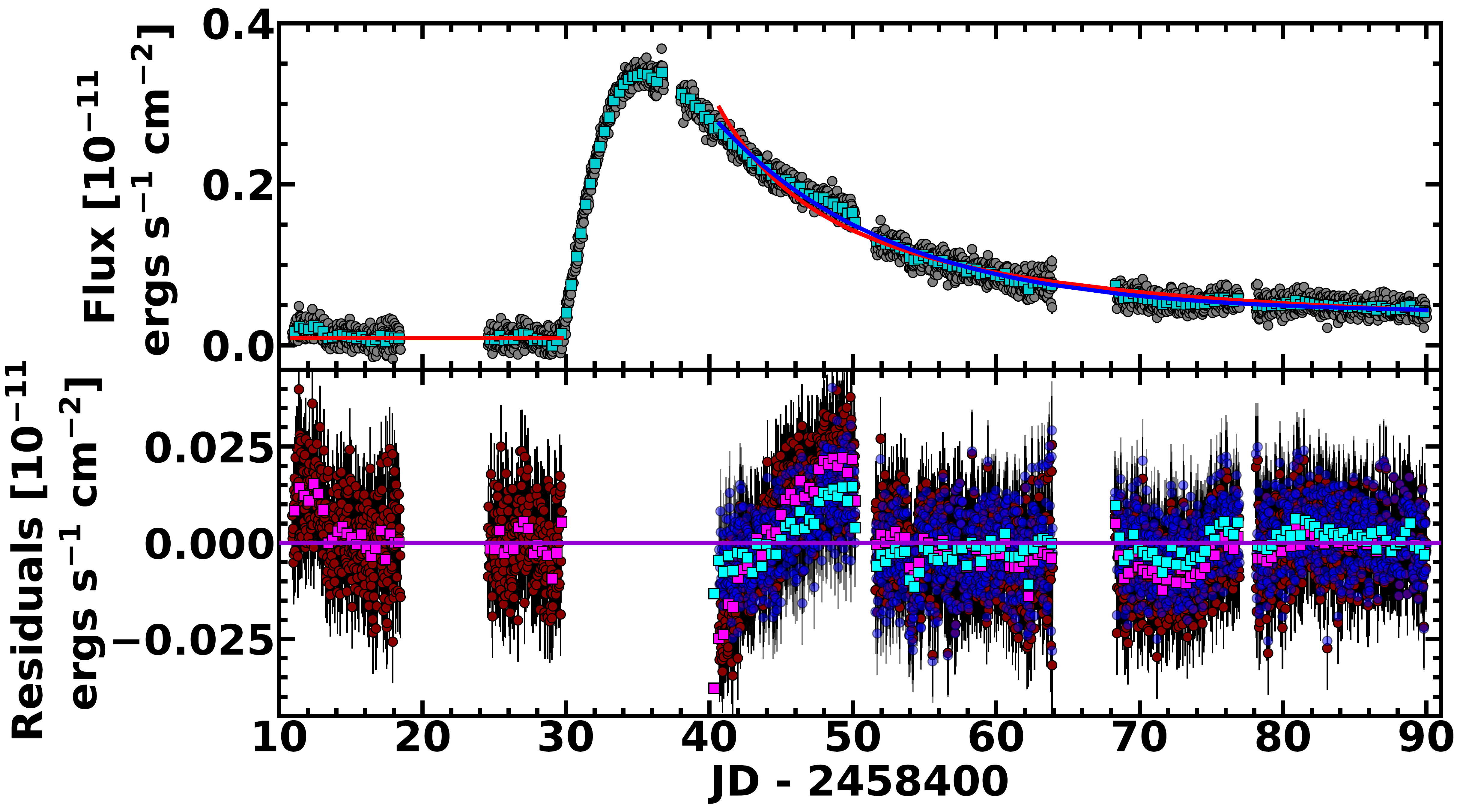}
    \caption{The declining phase models of the \textit{TESS} light curve starting 5 days past peak. The best-fit power-law decline is shown in red and the best-fit exponential decline is shown in blue. The \textit{TESS} data binned in 8 hour intervals are shown by turquoise squares. The bottom panel shows the residuals of the fits color-coded by model. Magenta and cyan squares show the residuals of the power-law decline and exponential decline compared to the binned \textit{TESS} data, respectively.}
    \label{fig:tess_decline}
\end{figure*}

The \textit{TESS} light curve of ASASSN-14ko's November 2018 outburst provides a unique view of the outburst morphology. Because \textit{TESS} observed ASASSN-14ko over three consecutive sectors, the light curve captures the pre-outburst quiescence, the full rise to peak, and the full decline back to quiescence. We first characterized the early-time rise of ASASSN-14ko with a power law model of 
\begin{equation}
    f = z \ \mathrm{ when } \ t < t_1 \mathrm{, and} 
\end{equation}
\begin{equation}
    f = z + h \left( \frac{t - t_1}{\mathrm{days}} \right) ^{\alpha} \  \mathrm{when} \  t > t_1\mathrm{,}
\end{equation}
consisting of residual background $z$, the time of rise $t_1$, a flux scale $h$, and the power law index $\alpha$.  We use the package $\mathrm{SCIPY.OPTIMIZE.CURVE \_ FIT}$'s \citep{scipy2020} Trust Region Reflective method to obtain a best fit model with parameters $z = -0.01 \pm 0.001$ ergs s$^{-1}$ cm$^{-2}$, $h = 0.1 \pm 0.01 $ ergs s$^{-1}$ cm$^{-2}$, $t_1 = 2458429.7 \pm 0.05 $ JD, and $\alpha = 1.01 \pm 0.07$. This fit is shown as the red curve in Figure \ref{fig:tess_rise_0.4mjy}. For this fit we inflated the error bars in quadrature by 0.0086 ergs s$^{-1}$ cm$^{-2}$ in order to make the reduced $\chi^2$ of the fit unity for 160 degrees of freedom. The high photometric precision of \textit{TESS} shows that the early-time rise was smooth, and that the time to peak in the \textit{TESS} filter is $5.60 \pm 0.05$ days from $t_1$ when measuring the peak from the data binned at 8-hour intervals. 

\begin{table*}[htp!]
\centering
\caption{ Best-fit parameters for the power-law and exponential decline models of the \textit{TESS} light curve starting the fits for different numbers of days after peak. $z$ for the power-law decline was $0.01\pm0.001$ ergs s$^{-1}$ cm$^{-2}$ for all start times post-peak. }
\begin{tabular*}{0.85\textwidth}{l l l l l}
\toprule
 &  & \textbf{Power-Law Decline} &  & \\
\pbox{25mm}{ Fit Start Time\\from Peak (Days)} & \pbox{25mm}{$t_0$ (JD-2458400)} & \pbox{25mm}{$h$ (ergs s$^{-1}$ cm$^{-2}$)} & $\alpha$ & \pbox{25mm}{$\chi^2$ per dof} \\ 

\hline

5 & $29.70\pm0.45$  & $-5.60\pm0.64$ & $-1.24\pm0.03$ & 1.09 \\ 
10 & $29.70\pm0.93$  & $-10.00\pm2.21$ & $-1.42\pm0.05$ & 0.84 \\
15 & $29.70\pm2.70$  & $-7.55\pm3.77$ & $-1.35\pm0.11$ & 0.76 \\ 
20 & $29.70\pm5.64$  & $-6.06\pm5.49$ & $-1.29\pm0.20$ & 0.76 \\ 

\hline 

 &  & \textbf{Exponential Decline} & & \\
\pbox{25mm}{ Fit Start Time\\from Peak (Days)} &  \pbox{25mm}{$a$ (ergs s$^{-1}$ cm$^{-2}$)} & \pbox{25mm}{$\tau$ (days)} & \pbox{25mm}{$c$ (days)} & \pbox{25mm}{$\chi^2$ per dof} \\

5 & $0.24\pm0.001$ & $12.22\pm0.11$ & $0.04\pm0.001$ & 0.77 \\
10 & $0.17\pm0.001$ & $10.35\pm0.13$ & $0.04\pm0.001$ & 0.70 \\
15 & $0.10\pm0.001$ & $11.40\pm0.32$ & $0.04\pm0.001$ & 0.65 \\
20 & $0.06\pm0.001$ & $10.79\pm0.49$ & $0.04\pm0.001$ & 0.64 \\ 
\hline 

\end{tabular*}
\label{tab:tess_decline_fit_params}
\end{table*}

Three TDEs have estimates of $\alpha$, all of which are steeper. The first TDE detected by \textit{TESS} was ASASSN-19bt, which had a power law index of $\alpha = 2.10 \pm 0.12$ \citep{holoien19}. ASASSN-19dj had $\alpha = 1.9 \pm 0.4$ using the ASAS-SN light curve \citep{hinkle202019dj}. Values of $\alpha$ $\sim$ 2 are similar to the fireball model used for the early-time evolution of SNe (\citealt{riess99}, \citealt{nugent11}). The third example, AT2019qiz, had a still steeper power-law index rise of $\alpha = 2.8 \pm 0.3$ \citep{nicholl2020}.

The photometric precision of the \textit{TESS} light curve also reveals that the decline was also remarkably smooth, as shown in Figure \ref{fig:tess_decline}. Since the rate of the decline in partial TDEs is predicted to be steeper than the canonical $t^{-5/3}$ model, we fit the \textit{TESS} light curve starting from five days after peak as 
\begin{equation}
    f = z - h\left( \frac{t - t_{\mathrm{0}}}{\mathrm{days}} \right)^{\alpha}
\end{equation}
with $t_{\mathrm{0}}$ being the time of disruption, constrained to be before the start of the rise, $t_1$, determined above. We again inflated the error bars in quadrature by 0.01 ergs s$^{-1}$ cm$^{-2}$ in order to make the reduced $\chi^2$ unity for 2611 degrees of freedom. The best-fit power-law has the parameters $t_0 = 2458429.70 \pm 0.45$ JD, which corresponds to the start of the rise, $z = 0.01 \pm 0.001$ ergs s$^{-1}$ cm$^{-2}$, $h = -5.60 \pm 0.64$ ergs s$^{-1}$ cm$^{-2}$, and $\alpha = -1.24 \pm 0.03$. As seen in Figure \ref{fig:tess_decline}, this smooth power law describes the decline well. The exponent, far from being steeper than the canonical $\alpha = -5/3 = -1.66$, is actually shallower.

We also fit the decline as an exponential decay of the form
\begin{equation}
    f = a e^{- (t - t_{\mathrm{peak}} ) / \tau}  + c
\end{equation}
which returns the best-fit parameters $a = 0.24 \pm 0.001$ ergs s$^{-1}$ cm$^{-2}$, $\tau = 12.22 \pm 0.11$ days, and $c = 0.04 \pm 0.001$ days. We set $t_{peak}$ to the peak of the \textit{TESS} light curve since its value is degenerate with the other parameters. This model has a reduced $\chi^2$ value of 0.77 for the same uncertainties. The exponential decline is, therefore, a better representation of the decline than the power-law decay model. Since starting the fit five days after peak is somewhat arbitrary, Table \ref{tab:tess_decline_fit_params} gives the results for fits starting 5, 10, 15, and 20 days after peak all for the same error model. The fit parameters are relatively stable and the exponential model is always the better fit.

\subsection{May 2020 Outburst}

The May 2020 outburst was the first to be predicted in advance, and it occurred as expected. We initially predicted the outburst to peak on MJD 58990.2 $\pm$ 1.9 based on our preliminary period estimate using the periodogram. The ASAS-SN $g$-band light curve actually peaked on MJD 58987.5$^{+2.5}_{-0.8}$, as shown in the last panel of Figure \ref{fig:ASASSN_LC}, consistent with the prediction. 

\begin{figure*}
    \centering
    \includegraphics[width=0.85\textwidth]{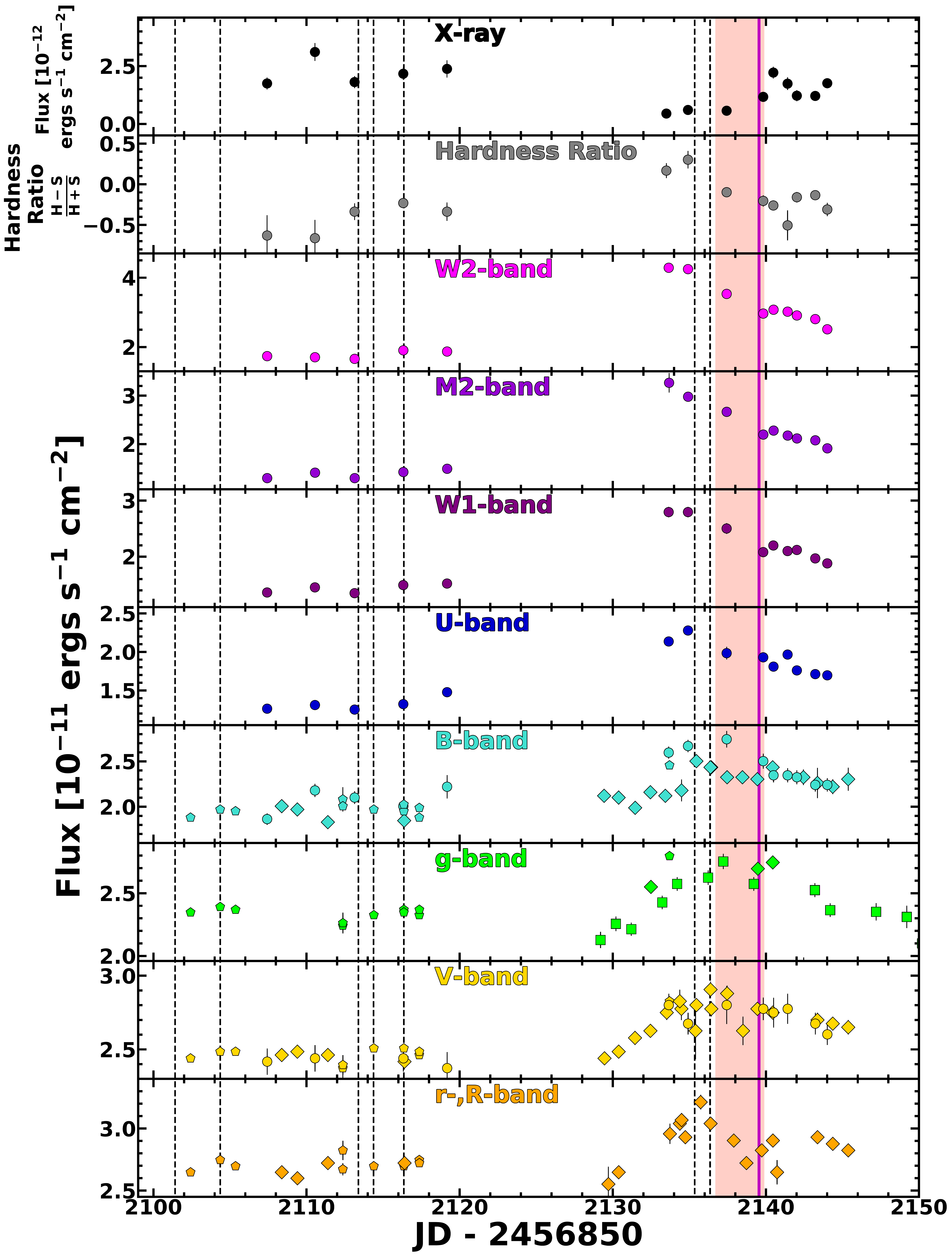}
    \caption{ Non-host subtracted photometry of the May 2020 outburst. ASASSN-14ko $g$-band photometry is represented as squares, \textit{Swift} data as circles, LCOGT data as pentagons, and data from amateur observatories as diamonds. $B$-band, $g$-band, and $V$-band data taken from different telescopes on the same day were averaged. \textit{Swift} $B$- and $V$-band data were converted to Johnson $B$ and $V$ magnitudes before being converted to flux to enable direct comparison with the ground-based data. Error bars are plotted but are frequently smaller than the size of the points. The red shaded region denotes the time of the ASAS-SN $g$-band peak on JD $2458987.5^{+2.4}_{-0.8}$ and the magenta vertical solid line shows the predicted peak for our model with a period derivative, as was also shown in Figure \ref{fig:ASASSN_LC}. The vertical dashed lines indicate the observation times of the LCOGT spectra shown in Figure \ref{fig:lcogt_spectra}. }
    \label{fig:nonhost_phot}
\end{figure*}

We requested Swift observations to monitor the outburst along with the LCOGT and amateur ground-based data. The non-host-subtracted light curves along with the X-ray hardness ratios are shown in Figure \ref{fig:nonhost_phot}. Unfortunately, ground-based observations were severely impacted by COVID-19 closures and the impending Sun constraint. The ASAS-SN light curve in the month leading up to the May 2020 outburst had a gap due to these closures. Observations were collected by the amateur astronomers, but large gaps exist in those light curve due to weather closures. The combined data set was processed as uniformly as possible, but the scatter prevents unambiguous interpretations of the multi-band light curve. 

\begin{figure}[th]
    \centering
    \includegraphics[width=\linewidth]{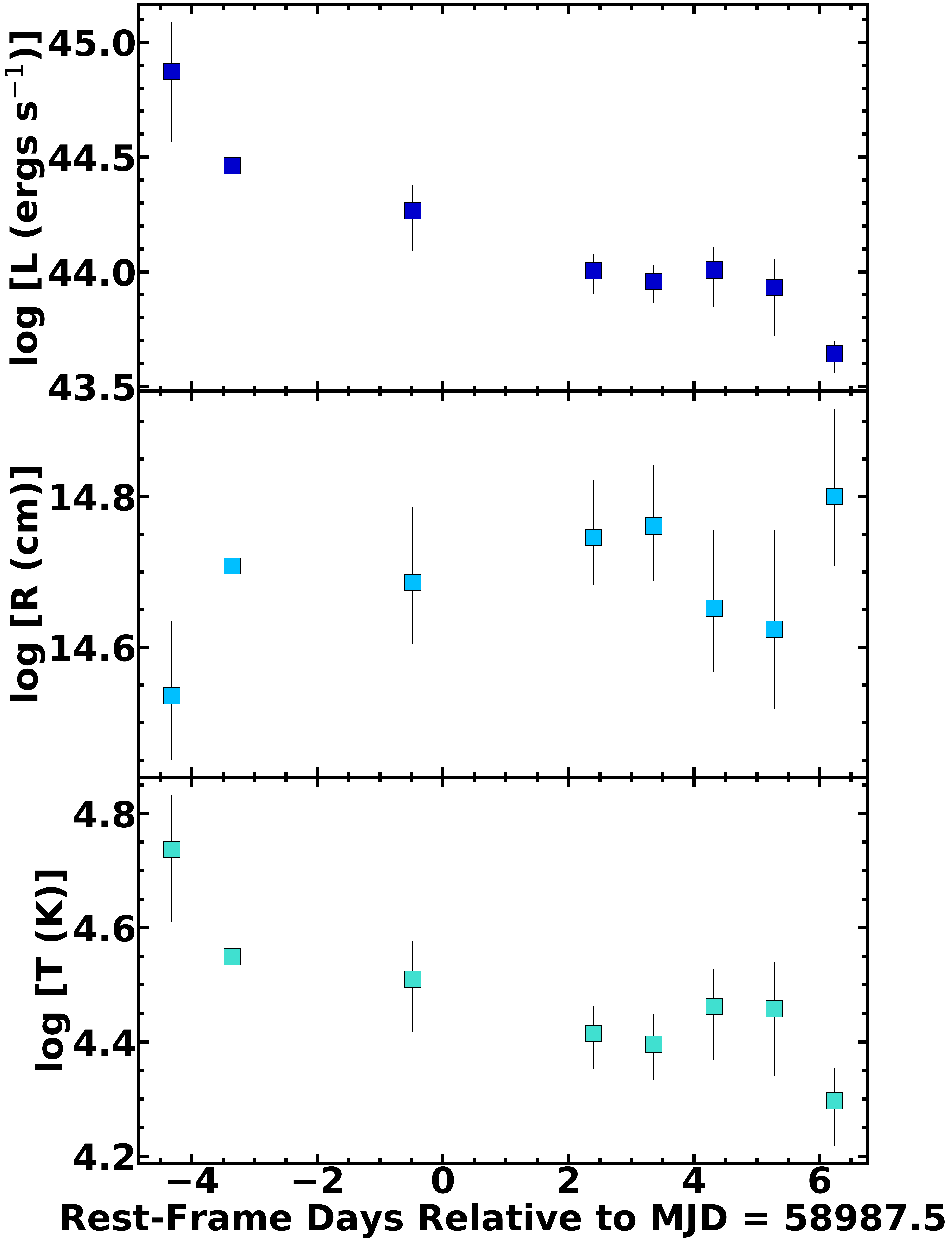}
    \caption{Evolution of the UV/optical blackbody luminosity (top panel), radius (middle panel), and temperature (bottom panel) during the May 2020 outburst based on the host-subtracted \textit{Swift} data and shown in Table \ref{tab:bb_vals}. The time is relative to the $g$-band peak on MJD 58987.5. The luminosity and temperature peak occurred several days prior to the $g$-band peak. }
    \label{fig:bbody_evol}
\end{figure}

The most interesting feature of Figure \ref{fig:nonhost_phot} is that we clearly see a wavelength dependence to the flux peak. While the rise was fully observed for the $B$-band and longer wavelengths, the $UV$ and $U$-band data clearly peaked at still earlier times and the optical bands lagged the $UV$-bands by several days. This is also apparent in the blackbody fits to the host-subtracted \textit{Swift} data shown in Figure \ref{fig:bbody_evol} and Table \ref{tab:bb_vals}. The peak blackbody luminosity and temperature occurred approximately four days prior to the ASAS-SN $g$-band peak. However, the blackbody radius remained roughly consistent over time.

The Swift XRT X-ray light curve did not follow the same trend. At peak optical flux, the X-ray flux had dropped by a factor of $\sim$4 and the spectrum became harder. The hardening of the emission could mean that the effective radius for the X-ray emission is shrinking, that a harder power-law flux component is increasing, or there is additional obscuration that is attenuating the soft X-ray photons. However, during the outburst peak, the hardness ratio decreased, making it difficult to reconcile the X-ray evolution with the UV/optical. 

\begin{figure}
    \centering
    \includegraphics[width=\linewidth]{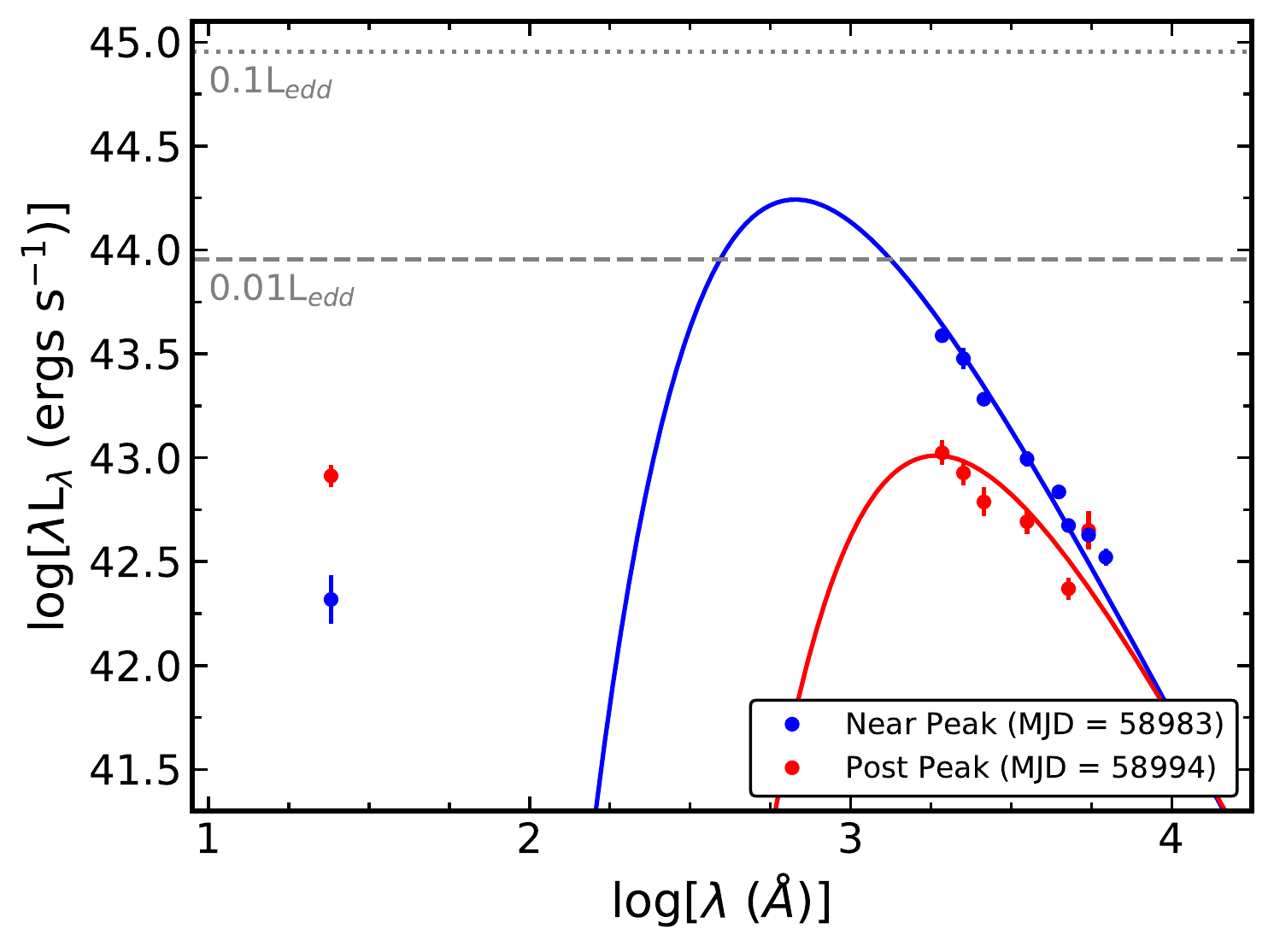}
    \caption{Host-subtracted spectral energy distribution at two epochs during the May 2020 outburst. The first epoch on MJD 58983 (blue) is four days prior to the $g$-band peak on MJD 58987.5. The second epoch on MJD 58994 (red) is seven days after the $g$-band peak. For the UV/optical, the data are shown as points while the lines represent the best-fit blackbody components. The X-ray luminosities are shown at the center of the 0.3-10 keV band over which the luminosity is determined. The dashed gray and dotted gray lines are 1\% and 10\% respectively of the Eddington luminosity for an SMBH of mass $7.2 \times 10^7 M\odot$.}
    \label{fig:sed}
\end{figure}

The UV/optical and X-ray SEDs at two epochs during the May 2020 outburst are shown in Figure \ref{fig:sed}. The two epochs correspond to four days before and seven days after the ASAS-SN $g$-band peak, which approximately corresponds to the $UV$ peak and post-maximum $UV$ quiescence. The peak luminosity is roughly 1\% of the Eddington luminosity derived in Section \ref{bhmass}.

\subsection{Comparison to TDEs}

\begin{figure}[t]
    \centering
    \includegraphics[width=\linewidth]{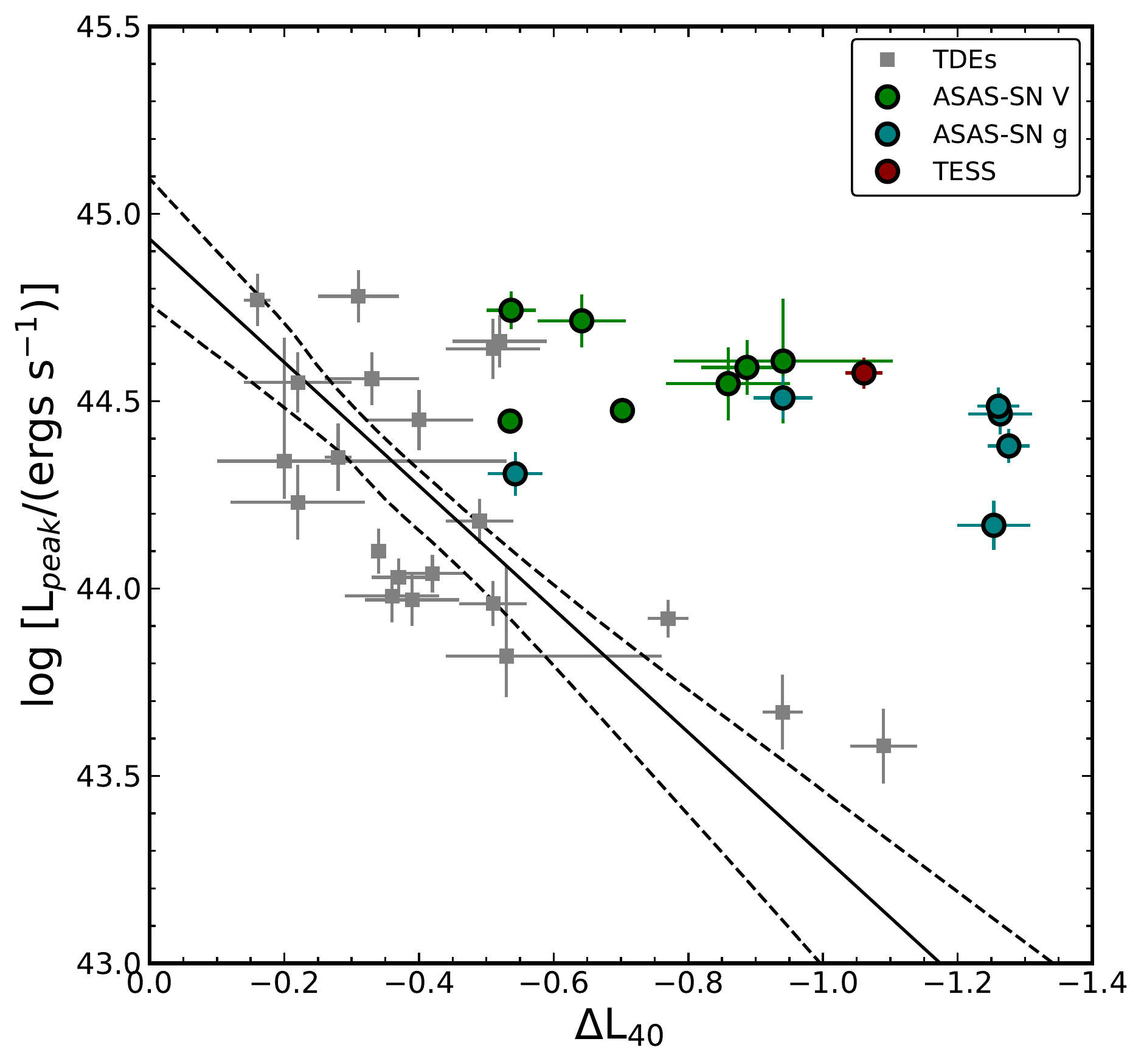} 
    \caption{The peak luminosity-decline rate for ASASSN-14ko's outbursts compared to twenty-one previously studied TDEs. Individual ASAS-SN $V$ (green circles), ASAS-SN $g$ (teal circles), and \textit{TESS} (red circle) epochs are shown along with the TDEs analyzed by \citet{hinkle20a} represented as gray squares. The black solid line is the best-fit line for the TDEs and the dashed black lines are the allowed range of uncertainty from the best fit line.}
    \label{fig:hinkle_relationl}
\end{figure}

We first compared each of ASASSN-14ko's outbursts to the peak luminosity-decline rate relation for previously-studied TDEs \citep{hinkle20a}. This relation describes a correlation between the peak luminosity and its decline luminosity over 40 days. First, we bolometrically corrected the ASAS-SN $V$- and $g$-band, and \textit{TESS} photometry using a temperature of 28,800 K, which is the median temperature of the most recent outburst based on the \textit{Swift} data. Then, following the procedure of \citet{hinkle20a}, we calculated a peak luminosity and the decline rate over 40 days for each outburst. In Figure \ref{fig:hinkle_relationl}, we compare the peak luminosity and decline rate of these outbursts to known TDEs. Even though the power-law slope of the decline is shallower than $t^{-5/3}$ (Section \ref{tess_decline_rise}), the actual decline rate is steeper than all TDEs of similar luminosity.

\begin{figure*}[t]
    \centering
    \includegraphics[width=0.5\linewidth]{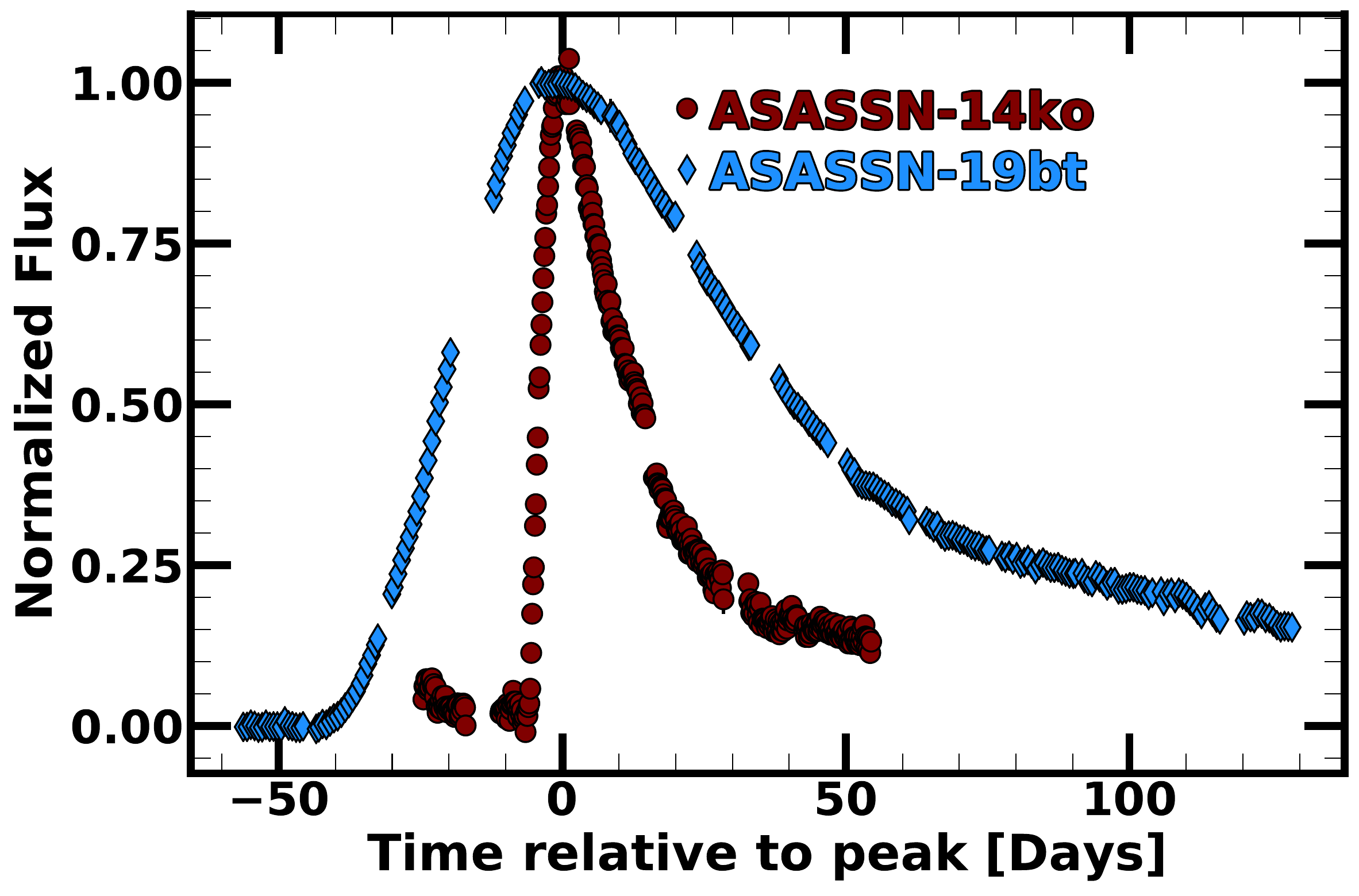}\hfill
    \includegraphics[width=0.5\linewidth]{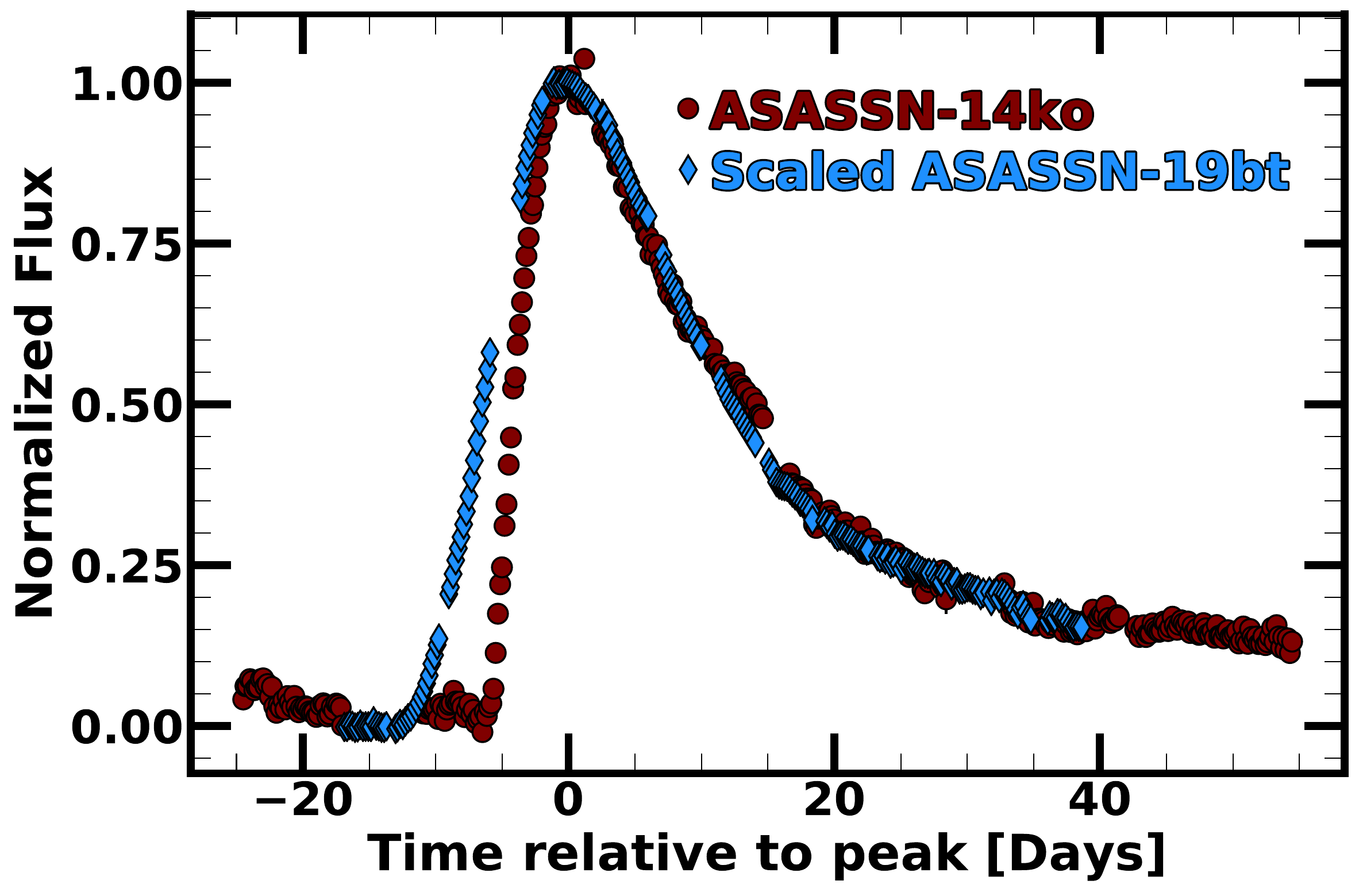}
    \caption{The \textit{TESS} light curves of ASASSN-14ko (red circles) binned in 4 hour intervals and ASASSN-19bt (blue diamonds) binned in 16 hour intervals. The left panel shows the light curves as observed, and the right panel shows the ASASSN-19bt light curve compressed in time by 30\% to align the light curve declines. }
    \label{fig:comp19bt}
\end{figure*}

We directly compare the \textit{TESS} light curves of ASASSN-14ko and ASASSN-19bt \citep{holoien19} in Figure \ref{fig:comp19bt}. ASASSN-14ko clearly evolves much more rapidly than ASASSN-19bt, with both a more rapid rise and a more rapid decline. However, the overall morphologies of the declining light curves are very similar if we compress the time relative to peak of ASASSN-19bt by 30\% in order to align the light curve declines as shown in the right panel of Figure \ref{fig:comp19bt}.

\section{Analysis of Spectra} \label{spectraanalysis}

\subsection{Evolution of the Optical Spectra} 

\begin{figure*}[t]
    \centering
    \includegraphics[width=0.7\linewidth]{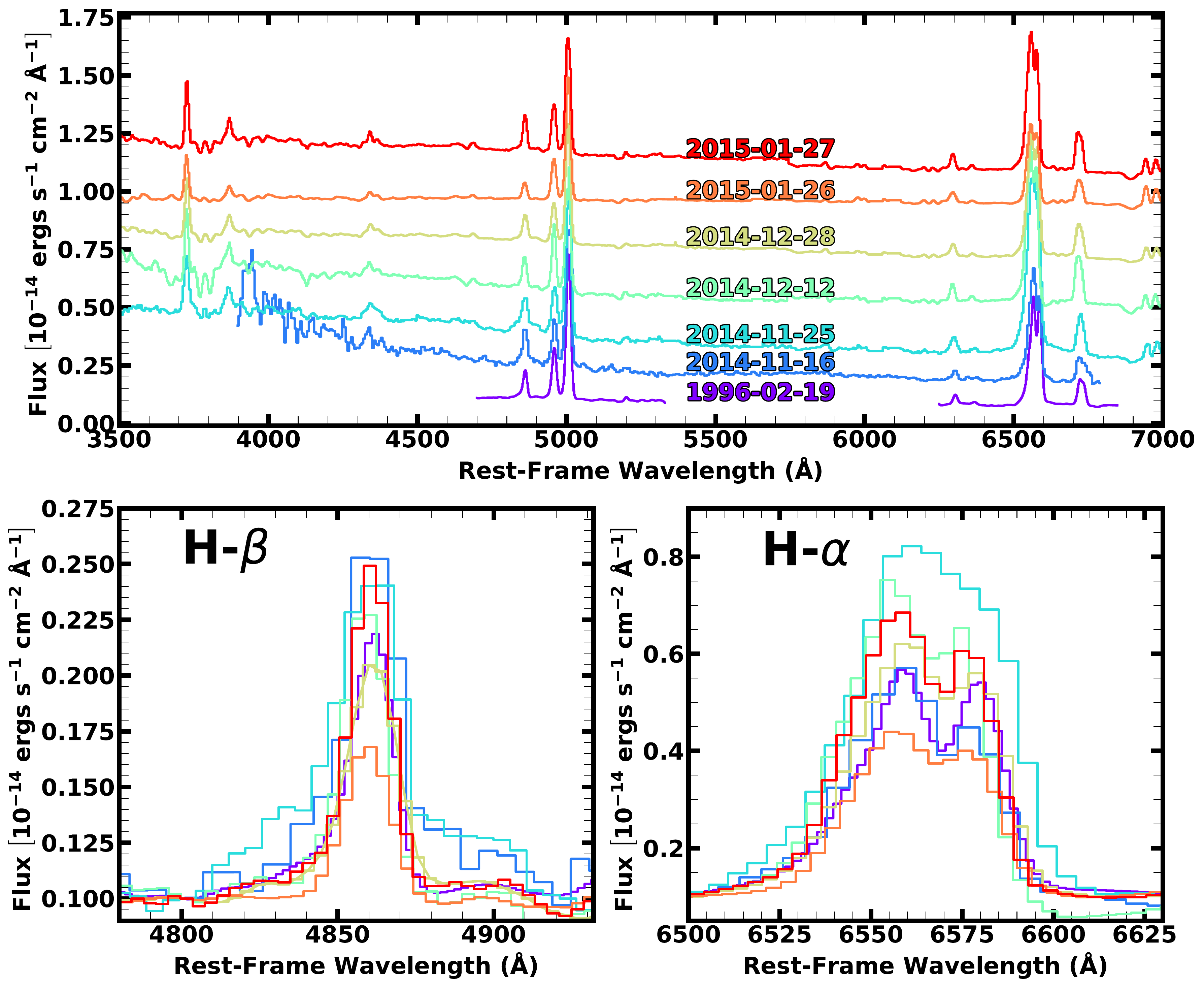}
    \caption{Spectra of \gal{} showing the change in the Balmer emission-line profiles. The earliest epoch from 1996-02-19 \citep{kewley01} is shown in violet. Subsequent spectra were taken after the discovery of ASASSN-14ko including the spectrum reported in \citet{holoien14ATELc} and five epochs from PESSTO. The spectra have been scaled using \texttt{mapspec} to put all of them onto a common flux scale. }
    \label{fig:archival_spectra}

    \centering
    \includegraphics[width=0.7\linewidth]{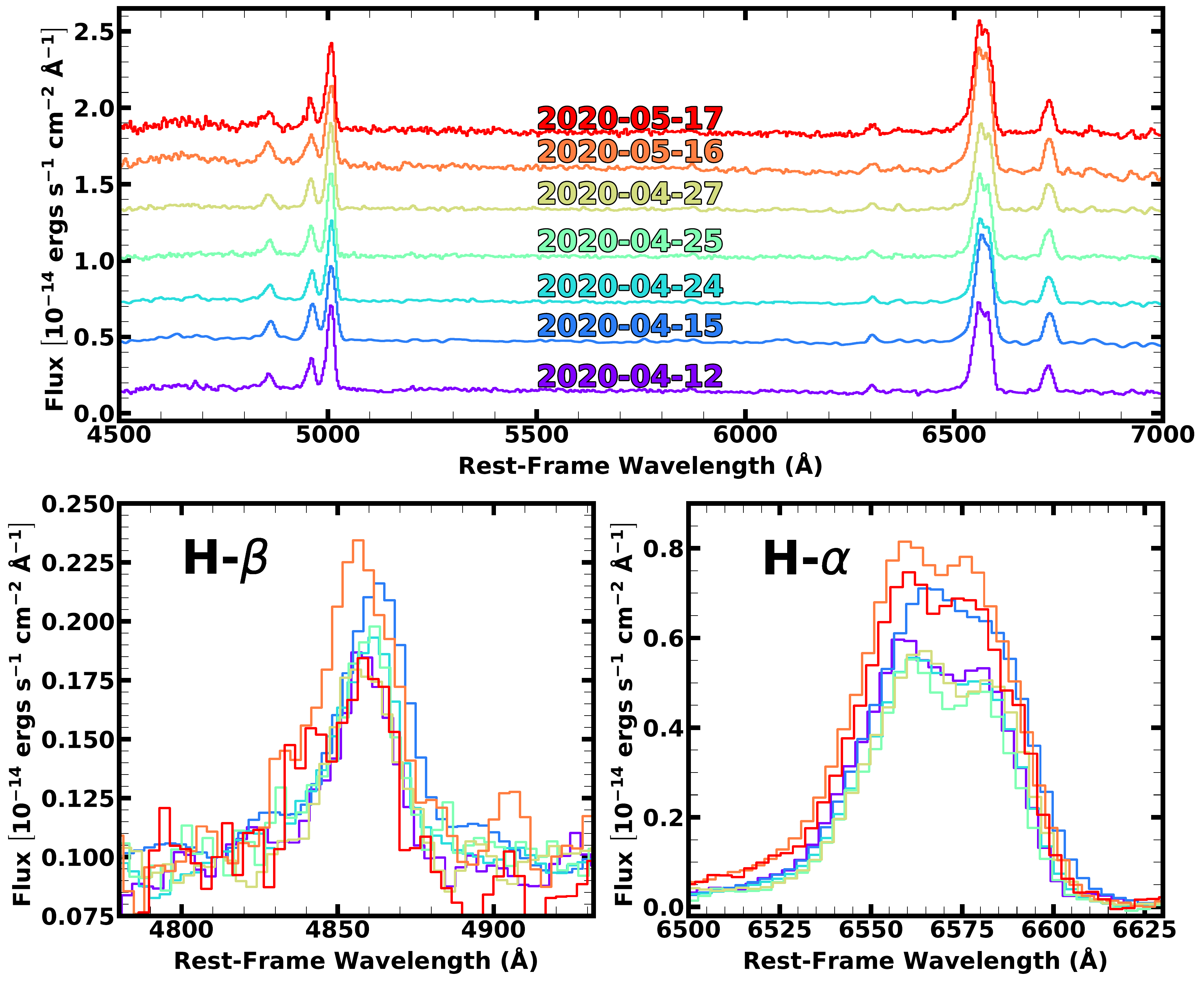}
    \caption{Spectra of \gal{} observed with the FLOYDS spectrograph at LCOGT during quiescence in April 2020 and during the May 2020 outburst. The first five spectra were obtained during quiescence and the last two spectra were obtained during the optical rise of May 2020 outburst. The spectra have been scaled using \texttt{mapspec} to put all of them onto a common flux scale.}
    \label{fig:lcogt_spectra}
\end{figure*}

We used optical spectra observed at different points in time and shown in Figure \ref{fig:archival_spectra} to first classify ASASSN-14ko using standard BPT diagnostics (\citealt{baldwin81}; \citealt{veilleux87}; \citealt{kauffmann03}; \citealt{kewley01}; \citealt{kewley06}). The 1996 spectrum and a weighted average of the 2014-2015 spectra both have line ratios consistent with an AGN. The measured ratios are given in Table \ref{tab:emline_bpt_table}. 

Next we examined the evolution of the H$\beta$ and H$\alpha$ emission-line profiles. The spectra from 2014-2015 were taken after ASASSN-14ko's optical peak, as indicated by orange tick marks in Figure \ref{fig:ASASSN_LC}. There is a noticeable change in the the Balmer line strengths and profiles during outburst. Since the line changes associated with the transient are at the redshift of \gal{}, we can be certain that these transients are not Galactic in origin. Directly comparing these spectra should be done with caution because the data were taken with different instruments and long-slit setups. However, it is apparent that the emission-line profile shapes clearly changed during the 2014 November outburst.

We obtained five spectra with the LCOGT FLOYDS spectrograph during quiescence in April 2020 and two spectra during the optical rise of the May 2020 outburst. We compared them to determine any change in the emission-line profiles as shown in Figure \ref{fig:lcogt_spectra}. Due to observatory airmass constraints and ASASSN-14ko's low position near the horizon from Siding Springs Observatory at that time, observations were not possible past 2020-05-17. These two observations occur one and two days before the optical peak, respectively, which coincides with the start of the decline in the UV, as shown by the vertical dashed lines in Figure \ref{fig:nonhost_phot} and the orange tick marks in Figure \ref{fig:ASASSN_LC}. A noticeable feature of the spectra during outburst is a blue wing around H$\beta$ near 4830\AA{} that is not present in the spectra taken during quiescence. This feature is similar to the broadened wings present in the 2014-11-16 and 2014-25 spectra which were taken during an outburst optical decline. This gives further evidence that the Balmer lines change during the outburst. 

\begin{table}[t]
\centering
\caption{Blackbody luminosity, radius, and temperature during the May 2020 outburst derived from the host-subtracted \textit{Swift} data.}
\begin{tabular*}{0.98\columnwidth}{l l l l}
\toprule

MJD & log [L (ergs s$^{-1}$)] & log [R (cm)] & log[T (K)] \\

\hline

58983 & $44.87^{+0.31}_{-0.22}$ & $14.54^{+0.09}_{-0.10}$ & $4.74^{+0.13}_{-0.10}$ \\
58984 & $44.46^{+0.12}_{-0.09}$ & $14.71^{+0.05}_{-0.06}$ & $4.55^{+0.06}_{-0.05}$ \\
58987 & $44.27^{+0.18}_{-0.11}$ & $14.69^{+0.08}_{-0.10}$ & $4.51^{+0.09}_{-0.07}$ \\
58990 & $44.00^{+0.10}_{-0.07}$ & $14.75^{+0.06}_{-0.08}$ & $4.42^{+0.06}_{-0.05}$ \\
58991 & $43.96^{+0.09}_{-0.07}$ & $14.76^{+0.07}_{-0.08}$ & $4.40^{+0.06}_{-0.05}$ \\
58992 & $44.01^{+0.16}_{-0.10}$ & $14.65^{+0.08}_{-0.10}$ & $4.46^{+0.09}_{-0.06}$ \\
58993 & $43.93^{+0.21}_{-0.12}$ & $14.62^{+0.11}_{-0.13}$ & $4.46^{+0.12}_{-0.08}$ \\
58994 & $43.64^{+0.09}_{-0.05}$ & $14.80^{+0.09}_{-0.12}$ & $4.30^{+0.08}_{-0.06}$ \\

\hline 

\end{tabular*}
\label{tab:bb_vals}
\end{table}

The similarities between the spectroscopic evolution of the two outbursts separated by nearly six years indicates that the spectroscopic change during the flare may be a consistent component of these events. The photometric and spectroscopic evolution appear closely tied, and higher temporal resolution of upcoming outbursts may reveal further connections between the rise and decline of the photometric light curves with morphological changes in the Balmer lines.

\begin{table}[t]
\centering
\caption{ Diagnostic emission line ratios log$_{10}$([O III]/H$\beta$), log$_{10}$([N II]/H$\alpha$), log$_{10}$([S II]/H$\alpha$), and log$_{10}$([O I]/H$\alpha$) used to distinguish AGN from H II-regions and classify AGN as either Seyferts or LINERs (\citealt{baldwin81}, \citealt{veilleux87}, \citealt{kewley01}, \citealt{kauffmann03}, \citealt{kewley06}). The 2014-2015 ratios were measured from a weighted average of the 2014-2015 spectra.} 
\begin{tabular*}{0.88\columnwidth}{l l l}
\toprule

Ratio Diagnostic & 1996 & 2014-2015  \\
\hline 
log$_{10}$([O III]/H$\beta$) & 0.73$\pm$0.04 & 0.81$\pm$ 0.12 \\
log$_{10}$([N II]/H$\alpha$) & $-$0.25$\pm$0.03 & $-$0.59$\pm$0.23 \\
log$_{10}$([S II]/H$\alpha$) & $-$0.57$\pm$0.03 & $-$0.61$\pm$0.07 \\
log$_{10}$([O I]/H$\alpha$) & $-$1.10$\pm$0.03 & $-$1.18$\pm$0.08 \\

\hline 

\end{tabular*}

\label{tab:emline_bpt_table}
\end{table}

\vspace{1cm}

\subsection{X-ray Analysis}
\begin{figure*}[ht]
    \centering
    \includegraphics[width=0.5\textwidth]{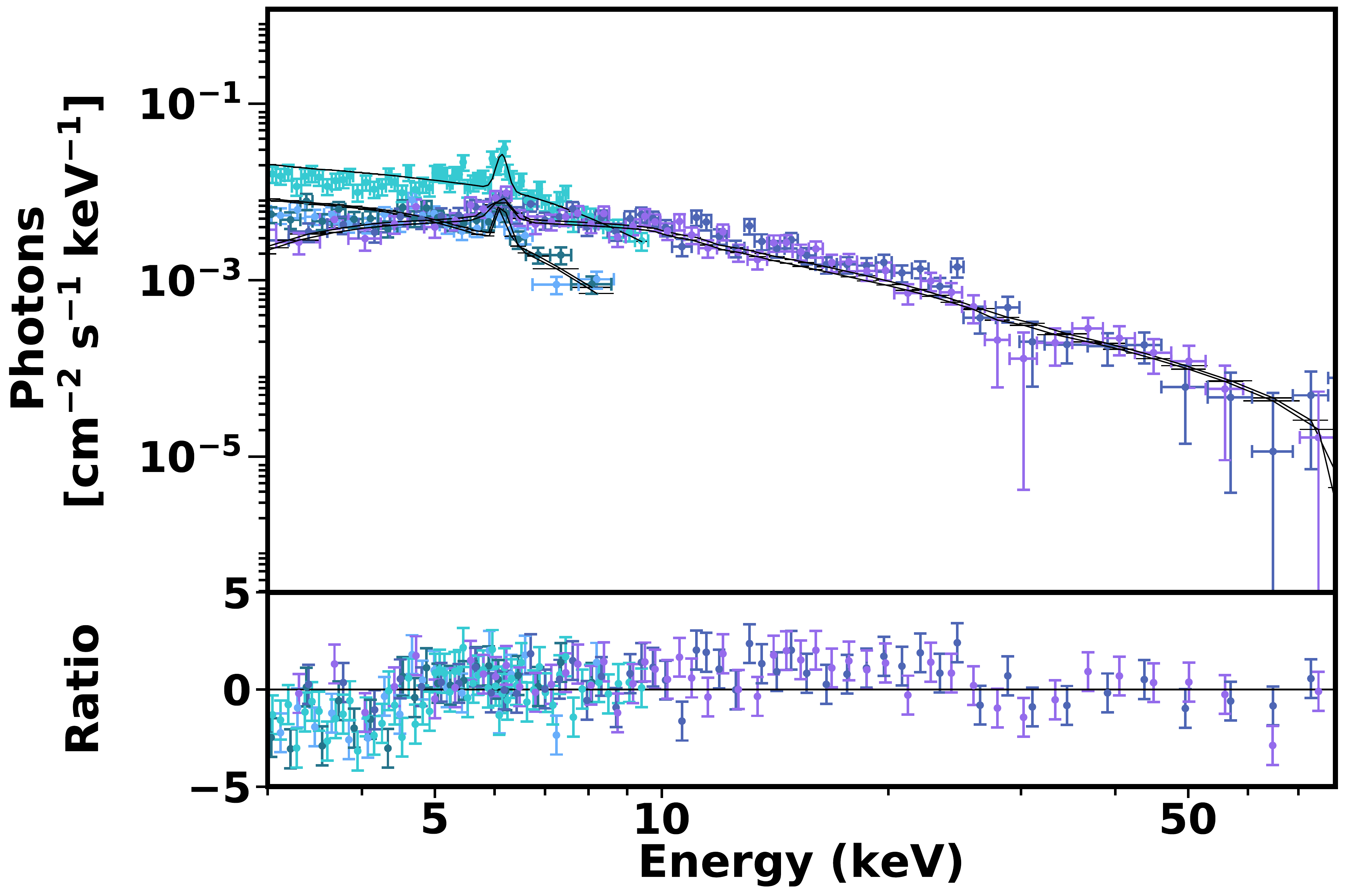}\hfill
    \includegraphics[width=0.5\textwidth]{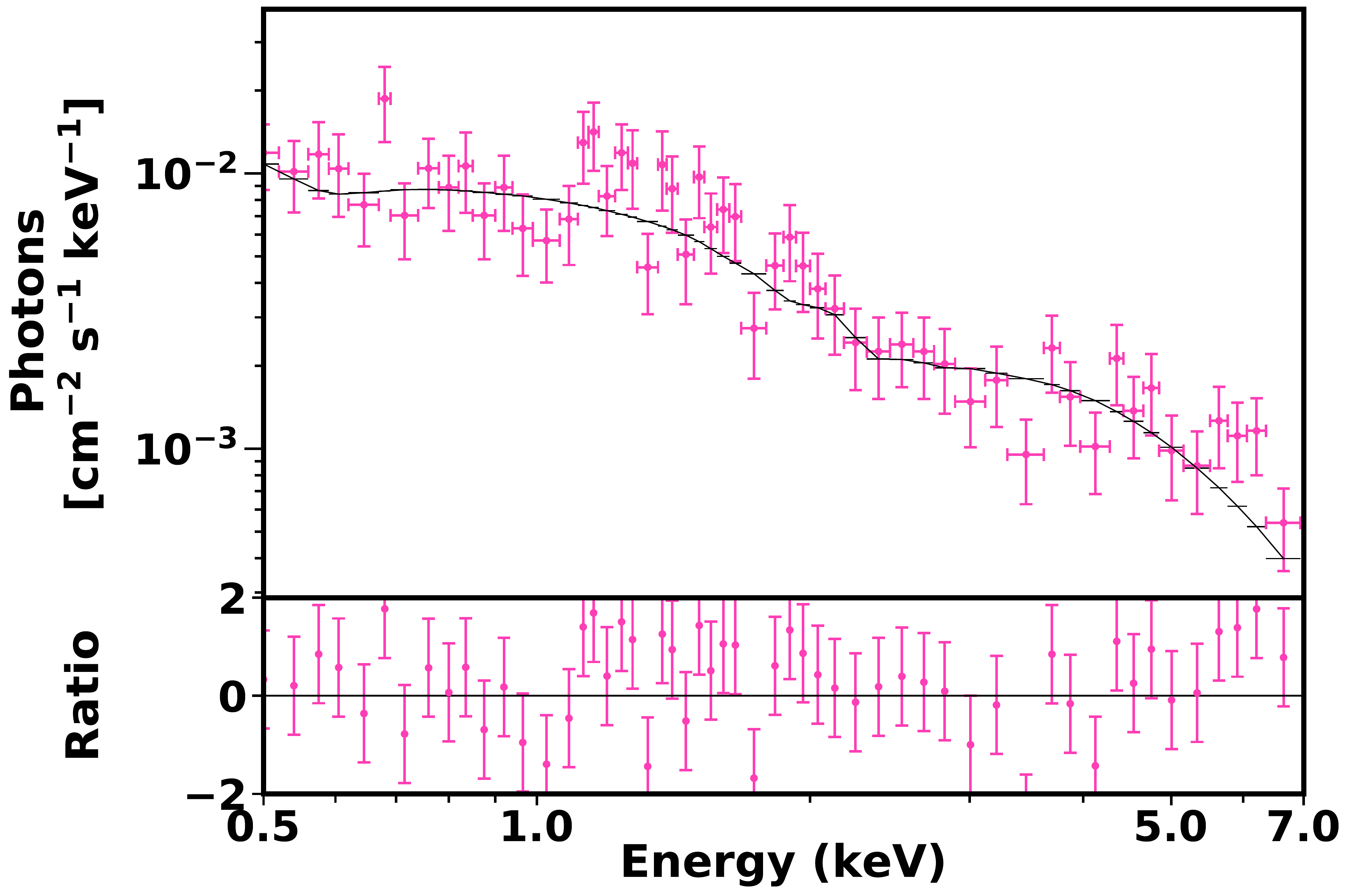}
    \caption{\textit{XMM-Newton}+\textit{NuSTAR} spectra (left) from MJD=57253.6 during a period of quiescence compared to the recent Swift spectra (right). The recent Swift observations do not show the Fe line around 6.4 keV, but this is likely due to the low signal-to-noise ratio.  }
    \label{fig:xray_spectrum}
\end{figure*}

We compared the two X-ray spectral epochs to characterize the X-ray emission evolution and are shown in Figure \ref{fig:xray_spectrum}. We modeled the 2015 \textit{XMM-Newton}+\textit{NuSTAR} spectra as a combination of a soft black body plus a power law. The black body model had a temperature of $0.15\pm0.01$ keV and the power law model had a $\Gamma = 0.87\pm0.04$. There is a strong 6.4 keV Fe line present that we include in the model as a Gaussian. The \textit{Swift} XRT spectrum was extracted from the merged data taken from $\sim$30 days prior to $\sim$9 days after the May 2020 flare, which corresponds to times during both quiescence and outburst. This spectrum was best fit by a power law with $\Gamma = 1.09^{+0.20}_{-0.22}$ and the fit was not improved by adding an additional black body with a best-fit temperature of $0.13\pm0.03$ keV. The \textit{Swift} spectrum had combined all the May 2020 observations because the signal-to-noise ratio of the spectrum would otherwise have been too low to assess its characteristics. With only $\sim$650 counts, poor statistics likely prevent any detection of the Fe line. The fit parameters are summarized in Table \ref{tab:xray_fit_params}.

The \textit{XMM-Newton} spectrum was taken during a period when the optical flare was quiescent, however the flux derived from the \textit{XMM-Newton} spectrum is similar to what we observed with \textit{Swift}. The similarities between the \textit{Swift} spectrum and the \textit{XMM-Newton} spectrum may indicate that even during times around the UV peak, the average properties of the X-ray emission during May 2020 may not have changed significantly from the properties of the deep \textit{XMM-Newton} observation. Future observations with better signal-to-noise will be essential to disentangle potential X-ray emission evolution between quiescence and outburst. There are nine epochs of \textit{XMM-Newton} slew observations of this galaxy observed between 2007-2019. However, given the sparse sampling and low signal to noise, there is no obvious pattern after phase folding these data. 

\section{Discussion} \label{discussion}
 Here we discuss several scenarios for ASASSN-14ko's periodic outbursts and some of their problems. We consider the possibilities that ASASSN-14ko is a sub-parsec SMBH binary system, a SMBH and perturbing massive star binary system, and, finally, a repeating partial TDE.  
 We do not consider stellar origins for the outbursts.  As noted earlier, a Galactic source is ruled out because the emission line changes associated with the outbursts occur at the redshift of the host.   At the redshift of the host, the peak luminosities are similar to those of luminous supernovae and so are too high for non-explosive stellar transients. The many quasi-periodic repetitions of similar luminosity rule out explosive possibilities. This appears to leave only phenomena associated with a central SMBH as possible explanations.

\subsection{ASASSN-14ko as a SMBH Binary System}
\label{sec:BinarySMBH}
As discussed in the introduction, apparently periodic emission from AGN is frequently interpreted as evidence for an SMBH binary.  However, there are three main inconsistencies with this scenario for ASASSN-14ko: (1) the period derivative; (2) the short life time of such a binary and (3) the lack of velocity shifts.  We will scale the results for black hole masses of $M_{BHa}=5 \times 10^7 \hat{M}_{BHa} M_\odot$, $M_{BHb} = 5 \times 10^7 \hat{M}_{BHb} M_\odot$ and a period of $P=114\hat{P}$~days.  

Due to the emission of gravitational waves, the binary will have a period derivative of 
\begin{equation}
   \dot{P} = 0.000056 \left[
   { f(e) \hat{M}_{BHa} \hat{M}_{BHb} \over \hat{P}^{8/3}
   \left(\hat{M}_{BHa}+\hat{M}_{BHb}\right)^{1/3}}\right],
     \label{eqn:pdot}
\end{equation}
where $f(e) = (1+73 e^2/24+37 e^4/96)(1-e^2)^{-7/2}$ is the dependence on the orbital ellipticity $e$, and a time to merger of 
\begin{equation}
   t_m = 2100 
    \left[{ g(e) \hat{P}^{8/3} \left(\hat{M}_{BHa}+\hat{M}_{BHb}\right)^{1/3}
       \over \hat{M}_{BHa} \hat{M}_{BHb}}\right] ~\hbox{years}
\end{equation}
where $g(e) = (1-e^2)^{7/2}$ \citep{peters64}.  The first problem is that the observed
period derivative is over an order of magnitude larger
than the scaling in Eqn.~\ref{eqn:pdot} for
black holes in a circular orbit.  This can be solved only by substantially increasing
the mass of at least one of the black holes, or by making the orbit significantly elliptical.
Both of these solutions to the mismatch in period derivatives will exacerbate the 
next two problems by reducing the binary lifetime and increasing the binary velocities. If the solution is to change the masses, the lines must be formed around the less massive SMBH, which also exacerbates the velocity problem.

The second problem is that finding a binary so close to merging is very improbable. The first way to phrase this is through what it requires for the properties of SMBH binaries in other galaxies.  The number density of $L_*$ 
galaxies is roughly $n \simeq 0.01h^3$~Mpc$^{-3}$ (e.g., \citealt{kochanek2001}), so there are roughly $N_g \simeq 3 \times 10^5$ such galaxies inside the distance to \gal{}.  To have one system merging in the next $t_0=1000$~years implies that all of these galaxies must contain binaries that will merge in the next $N_g t_0 \simeq 300$~million years. This in turn implies that they must all have orbital periods shorter than $\sim N_g^{3/8} P_0 \simeq 35$~years, which would seem to make the problem of finding binary SMBH systems rather trivial rather than being as difficult as it appears to be in practice.  The second way to phrase the problem is that if there is one SMBH binary with a period $<1/3$~year in this volume, there are $3^{8/3}=19$ with periods $<1$~year, 350 with periods $<3$~years, and 8700 with periods $<10$~years.

\begin{table*}[hbt!]
\centering
\caption{Best-fit parameters for the \textit{Swift} and \textit{XMM-Newton}+\textit{NuSTAR} X-ray spectra. Both models used a redshift of 0.0425 and a neutral hydrogen column density $N_H$ of $3.5 \times 10^{20}$ cm$^{-2}$ frozen to the Galactic column density \citep{HI4PI2016}. The \textit{XMM-Newton}+\textit{NuSTAR} spectrum included a Gaussian line energy of $6.39\pm0.03$ keV to capture the 6.4 keV Fe line. The \textit{Swift} fit had 56 degrees of freedom and the \textit{XMM-Newton}+\textit{NuSTAR} fit had 470 degrees of freedom.    }
\begin{tabular*}{0.70\textwidth}{l l l l l}
\toprule
 Instrument &  $kT$ (keV) & \pbox{30mm}{ Blackbody \\ Normalization (K)} &  Photon Index $\Gamma$ & $\chi^2$ per dof \\ 
\hline
\textit{Swift} & $0.13\pm0.03$ & $79.55_{-46.70}^{+137.60}$ & $1.09_{-0.22}^{+0.20}$ & $1.07$ \\
\textit{XMM+NuSTAR} & $0.15\pm0.01$ & $83.90_{-21.80}^{+32.30}$ & $0.87\pm0.04$ & 1.64 \\

\hline 
\end{tabular*}
\label{tab:xray_fit_params}
\end{table*}

The third problem for an SMBH binary is the high velocity scale, with
\begin{equation}
    v_a \simeq 16000 \left[ { \hat{M}_{BHa}+\hat{M}_{BHb} \over \hat{P} }\right]^{1/3} { \hat{M}_{BHb} \sin i \over \hat{M}_{BHa}+\hat{M}_{BHb} }~\hbox{km/s}
\end{equation}
for circular orbits.  While our phase sampling is poor, we arguably can set a limit on
any H$\beta$ line shifts of $<50$\AA, which corresponds to a velocity shift limit
of $<3000$~km/s.  If our estimate of $\dot{P}$ is correct, this problem cannot be
solved by reducing the masses or invoking a large mass ratio with the emission lines
being formed by material associated with the more massive SMBH.  Similarly, raising the 
ellipticity to solve the $\dot{P}$ problem makes this problem worse because of the higher 
velocities at pericenter compared to a circular orbit of the same period.  The velocity problem
can only be solved by making the system nearly face on or by relying on the poor spectral
sampling to hide the velocity shifts.

\subsection{ASASSN-14ko as a SMBH with a Perturbing Star}

Rather than a binary system with two SMBHs, the periodic outbursts could be driven by a star orbiting a single SMBH. Because stars are far more common than SMBHs and the gravitational wave merger lifetimes are now far longer, the probability argument against an SMBH binary is removed. The long gravitational wave merger time does mean that the observed $\dot{P}$ must have a different origin such as viscous interactions between the disk and the star, although estimating this effect is non-trivial.

\begin{figure}
    \centering
    \includegraphics[width=\linewidth]{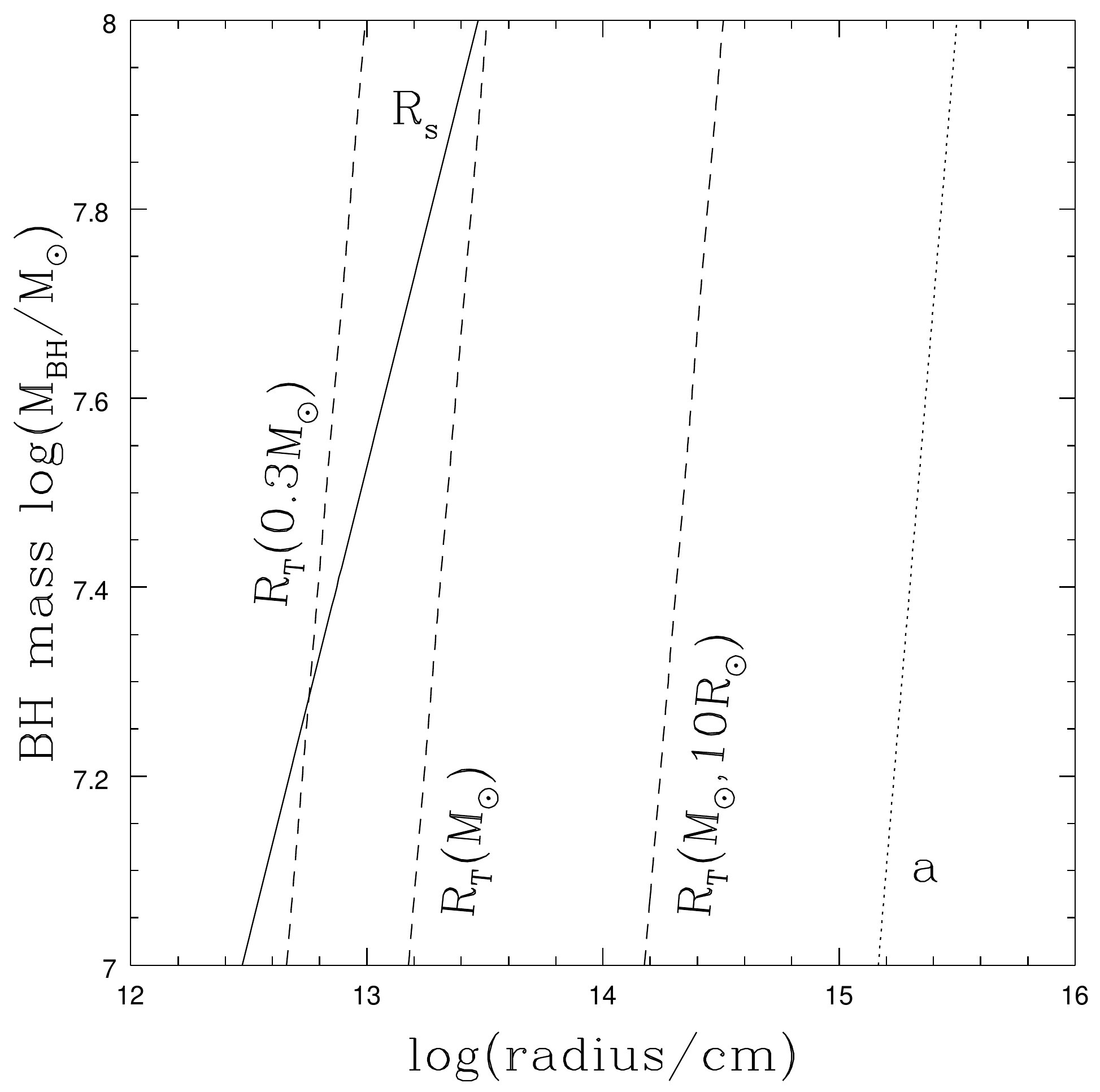}
    \caption{The Schwarzschild radius ($R_s$, solid), semimajor axis ($a$, dotted) and several tidal radii ($R_T$, dashed) as a function of SMBH mass $M_{BH}$. The tidal radii are for a $0.3M_\odot$ main-sequence star ($R_T(0.3M_\odot)$), the Sun ($R_T(1.0M_\odot)$), and a star with the mass of the Sun and a radius of $10R_\odot$ ($R_T(M_\odot,10R_\odot$).  On these logarithmic scales, the effects of spin on the BH horizon and the tidal limits are modest.}
    \label{fig:tidal}
\end{figure}

We now assume a star of mass $M_*=\hat{M}_*M_\odot$ and radius $R_*=\hat{R}_* R_\odot$ orbiting a black hole of mass $M_{BH}= 5\times 10^7 \hat{M}_{BH}M_\odot$. 
There are three length
scales of immediate interest, and we show their
relative values and their dependence on SMBH mass
in Figure \ref{fig:tidal}.  
The Schwarzschild radius of the black hole is
\begin{equation}
     R_s = 1.5\times 10^{13} \hat{M}_{BH}~\hbox{cm},
\end{equation}
the tidal (Roche) radius to disrupt the star is
\begin{equation}
    R_T = 2.5 \times 10^{13} 
     \hat{R}_* \hat{M}_*^{-1/3}
     \hat{M}_{BH}^{1/3}~\hbox{cm}
\end{equation}
and the orbital semi-major axis is
\begin{equation}
      a = 2.5 \times 10^{15}
     \hat{P}^{2/3}
    \hat{M}_{BH}^{1/3}~\hbox{cm}.
\end{equation}
Finally, for a pericentric radius $R_p$, the ellipticity
of the orbit is
\begin{equation}
      e \simeq 1 - 0.01 { R_p \over R_T}
           { \hat{R}_* \over \hat{M}_*^{1/3}}.
\end{equation}
For some star/SMBH scenarios, the true period would be twice the observed period, which would increase the semimajor axis by $0.2$~dex.

From Figure ~\ref{fig:tidal}, we see that there is no problem having a main sequence star orbiting an SMBH in this mass range without serious tidal effects provided the orbital ellipticity is moderate.  Particularly over many orbits, there would be some relativistic effects.  For example, the orbital precession per orbit is 
\begin{equation}
      {\dot{\omega} P \over 2 \pi}
      = { 6 \pi GM_{BH} \over a c^2 (1-e^2)}
      \simeq { 3 \over 4 } { R_s \over R_p} \mathrm{,}
\end{equation}
for a Schwarzschild black hole.  This means that we would expect systematic changes with time independent of the mechanism driving the flares.

\begin{figure*}[t]
    \centering
    \includegraphics[width=\linewidth]{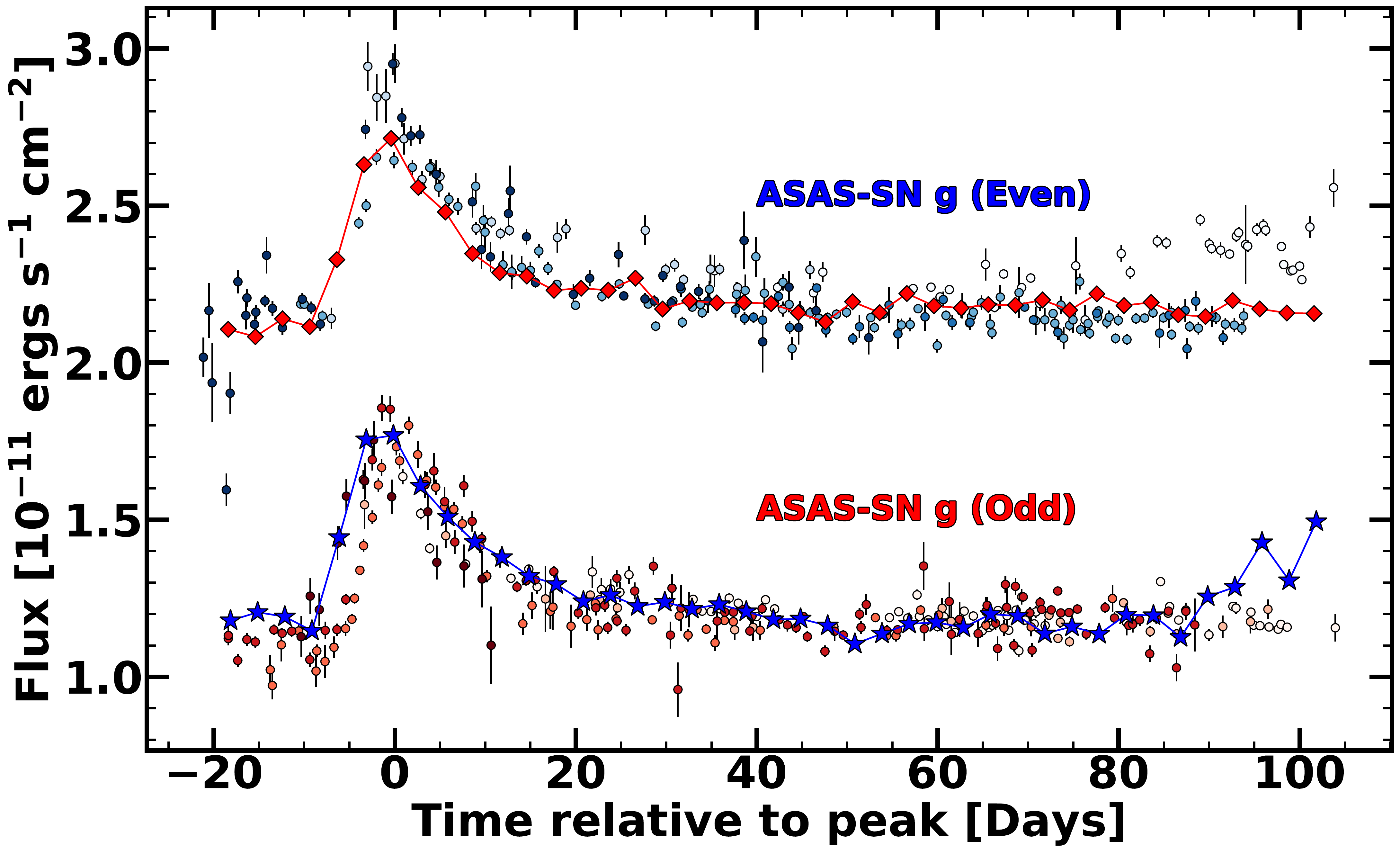}
    \caption{The stacked ASAS-SN $g$-band light curves separated by even (blue) and odd (red) outbursts as a function of phase. The light curves are offset for clarity and the data of each phase are given a different color shade. The binned even light curve, shown by star symbols, is superimposed over the odd light curves, and vice versa, with the binned odd light curve shown by diamond symbols. }
    \label{fig:phasefold_evenodd}
\end{figure*}

If the star is not being tidally disrupted, then the flares must be driven by periodic disturbances of the accretion flows as the star passes through the accretion disk.  However, a star simply embedded in the disk would represent a continuous perturbation that is unlikely to drive periodic flares.  Flaring would seem to require a stellar orbit at a significant inclination angle relative to the disk.   The star would then make two passages through the disk per orbit, so the orbital period would likely be twice the flare period.

Each orbit would produce a pair of flares with the spacing dependant on the orbital eccentricity and the argument of periapsis ($\omega$) relative to the accretion disk. The separation between pairs will increase as the orbit becomes more eccentric.  While eccentric orbits with $\omega$ near 0 or $\pi{}$ can have equally spaced encounters, the star will encounter the disk at different radii/temperatures. Given that the flare spacing, profiles, and amplitudes are all essentially constant between cycles, a perturbing star would seem be required to be on an inclined but nearly circular orbit.  Even then, the similarity of the flares seems odd because one would expect encounters with the star moving away from the observer and into the disk to differ from the reverse, except for nearly edge on viewing angles. In Figure \ref{fig:phasefold_evenodd} we see that the even and odd flare profiles are very similar in amplitude, shape, and duration which is difficult to reconcile with this model. 

Finally, while an inclined, circular orbit of a star orbiting a SMBH might be able to perturb the accretion disk with the right frequency, there is no obvious time scale in disks to then make the flares so short in duration.

\subsection{ASASSN-14ko as a Repeated Partial TDE}

A third possibility to explain ASASSN-14ko's periodic outbursts is as a repeating TDE that is partially disrupted after each passage close to the central SMBH.
While it requires fine tuning to have a main sequence star pass close to its tidal limit but remain outside the SMBH horizon, we can see from Figure ~\ref{fig:tidal} that it is relatively easy for an evolved star on an elliptical orbit to do so. As discussed in the introduction, giants are also the most likely candidates for partial disruptions that could power periodic flares.   \citet{macleod12} and \citet{guillochon13} find that the star will begin to lose mass once $\beta = R_T/R_p < 0.5$. 

The peak luminosities of the flares, $L_p \simeq 5 \times 10^{44}$~erg/s correspond to peak accretion rates of $\dot{M}_p \simeq 0.1\epsilon_{0.1}^{-1}M_\odot$/year where $\epsilon=0.1\epsilon_{0.1}$ is the accretion efficiency. If the peaks last $\sim 10$~days, the accreted mass of $\Delta M \simeq 0.003 \epsilon_{0.1}^{-1}M_\odot$ is certainly low enough to allow repeated outbursts on this scale.   Note, however, that these accretion rates are significantly higher than envisioned by \citet{macleod13} and the time scales are much shorter. 

\citet{ryu2020} found that the change in the orbital specific energy of the star in a partial disruption is $f \sim 10^{-3}$ of the specific energy scale $ G M_{BH} R_*/R_t^2$ of the stripped debris.  They find both positive and negative energy changes, so there is no prediction of the sign of the changes. The period derivative measured in Section \ref{pdot} implies a change in the orbital specific energy of $ G M_{BH} \dot{P}/3 a$.  This means that we should expect a period derivative of 
\begin{equation}
   | \dot{P}| = { 3 f a R_* \over R_t^2 } \simeq 0.8 f { \hat{M}_*^{2/3} \hat{P}^{2/3} \over
        \hat{M}_{BH}^{1/3} \hat{R}_*}.
\end{equation}
For $f \sim 10^{-3}$, this implies $|\dot{P}| \sim 10^{-3}$ with relatively little sensitivity to the exact values of the parameters and remarkably close to the measured period derivative.  The agreement is perhaps more remarkable because the \citet{ryu2020} simulations were for a single pericentric passage of a main sequence star on an initially parabolic orbit, rather different from the orbit required here.   However, the orbit of the puffy stellar merger remnant in Figure 13 (top) of \citet{antonini2011} has a semi-major axis shrinking as $\Delta a/a \sim 10^{-3}$ per orbit, which is the same order of magnitude.  The example in the lower panel of this figure shows very little orbital evolution but also shows very little ongoing mass loss.   
Note that the orbital changes essentially occur with the pericenter fixed because the tidal interactions are only important at pericenter.   Because the structure of the star must be changing with each pericentric passage due to the mass loss, torques and heating, $\dot{P}$ presumably cannot be constant on longer time scales.

The partial TDE hypothesis also seems better able to explain the similarity of the flares since each pericentric passage is almost identical in geometry to the previous and the required mass loss rates appear to be modest.  However, they cannot truly be identical since the orbital geometry must slowly change due to precession (relativistic and tidal) and the mass lost over tens of encounters ceases to be modest.  
Overall, the repeating, partial TDE interpretation seems most consistent with the available observations.   

\vspace{1cm}

\section{Summary}
Although ASASSN-14ko was first thought to be a supernova, the subsequent six years of ASAS-SN $V$- and $g$-band data show that the flares occur at regular intervals. The 17 flares observed to date are well modeled using a period of $P_0 = 114.2 \pm 0.4$ days and period derivative of $\dot{P} = -0.0017 \pm 0.0003$. Adopting this model, the next two flares will peak in the optical on UT $2020\text{-}09\text{-}7.4 \pm 1.1$ and UT $2020\text{-}12\text{-}26.5 \pm 1.4$. 

In addition to ASAS-SN, ATLAS, and \textit{Swift} multi-wavelength photometric data, \textit{TESS} observed ASASSN-14ko during its November 2018 outburst. The \textit{TESS} light curve has a decline rate that is dissimilar to previously studied TDEs and a rapid rise to peak occurring over $5.60 \pm 0.05$ days. The \textit{TESS} data also show that the rise and decline were smooth and lack short-timescale variability. The individual outbursts are morphologically very similar over the six year baseline of observations. While a host of problems interfered with studying the May 2020 outburst well, there was clear evidence that the outburst peaks a few days earlier in the UV than in the optical. Spectra taken during and prior to the May 2020 outburst revealed morphological changes around H$\beta$ during the flare which was similar to what occurred during the 2014 outburst. This suggests that morphological changes in the emission lines are consistently associated with the optical outbursts over time.  
We examined several possible scenarios to explain the cause of this AGN's unusual behavior, including the presence of a SMBH binary, a SMBH plus a perturbing massive star, or a repeating partial TDE. Between these scenarios, we favor a repeating partial TDE. We believe that any stellar transients or explosions whether Galactic or in the host are ruled out.  
The most important next step is to time and study the flares more closely across the electromagnetic spectrum. The relatively short period and system brightness make this relatively easy. ASASSN-14ko will be observed by \textit{TESS} again in Sectors 31-33 during the predicted December 2020 outburst. These observations will give further constraints on the nature of these outbursts, and presents a unique opportunity to do a detailed reverberation mapping analysis of the system.  

{\bf Software:}
ftools \citep{blackburn95}, \textit{NuSTAR} Data Analysis Software (v1.8.0)

\acknowledgments
We thank Chris Ashall, Patricia T. Boyd, Bradley Peterson, Richard Pogge, and Jennifer van Saders for helpful discussion. 
We thank the Las Cumbres Observatory and its staff for its continuing support of the ASAS-SN project. ASAS-SN is supported by the Gordon and Betty Moore Foundation through grant GBMF5490 to the Ohio State University, and NSF grants AST-1515927 and AST-1908570. Development of ASAS-SN has been supported by NSF grant AST-0908816, the Mt. Cuba Astronomical Foundation, the Center for Cosmology and AstroParticle Physics at the Ohio State University, the Chinese Academy of Sciences South America Center for Astronomy (CAS- SACA), and the Villum Foundation. 
AVP acknowledges support from the NASA Fellowship through grant 80NSSC19K1679. BJS, CSK, and KZS are supported by NSF grant AST-1907570. BJS is also supported by NASA grant 80NSSC19K1717 and NSF grants AST-1920392 and AST-1911074. CSK and KZS are supported by NSF grant AST-181440. KAA is supported by the Danish National Research Foundation (DNRF132).  MAT acknowledges support from the DOE CSGF through grant DE-SC0019323.  Support for JLP is provided in part by FONDECYT through the grant 1151445 and by the Ministry of Economy, Development, and Tourism's Millennium Science Initiative through grant IC120009, awarded to The Millennium Institute of Astrophysics, MAS. TAT is supported in part by Scialog Scholar grant 24215 from the Research Corporation.
Parts of this research were supported by the Australian Research Council Centre of Excellence for All Sky Astrophysics in 3 Dimensions (ASTRO 3D), through project number CE170100013.

\bibliographystyle{aasjournal}
\bibliography{references}

\end{document}